\pdfoutput=1
\documentclass[pdflatex,sn-basic]{sn-jnl}

\usepackage{graphicx}%
\usepackage{multirow}%
\usepackage{amsmath,amssymb,amsfonts}%
\usepackage{amsthm}%
\usepackage{mathrsfs}%
\DeclareFontFamily{U}{rsfs}{\skewchar\font127 }
\DeclareFontShape{U}{rsfs}{m}{n}{%
   <-6> rsfs5
   <6-8> rsfs7
   <8-> rsfs10
}{}
\usepackage[title]{appendix}%
\usepackage{xcolor}%
\usepackage{textcomp}%
\usepackage{manyfoot}%
\usepackage{booktabs}%
\usepackage{algorithm}%
\usepackage{algorithmicx}%
\usepackage{algpseudocode}%
\usepackage{listings}%
\usepackage{siunitx}%
\usepackage{tikz}

\usepackage{bookmark}

\definecolor{cRed}{HTML}{E41A1C}
\definecolor{cBlue}{HTML}{377EB8}
\definecolor{cGreen}{HTML}{4DAF4A}
\definecolor{cPurple}{HTML}{984EA3}
\definecolor{cOrange}{HTML}{FF7F00}
\definecolor{cBrown}{HTML}{A65628}
\definecolor{cPink}{HTML}{F781BF}
\definecolor{cTeal}{HTML}{1B9E77}
\definecolor{cRust}{HTML}{D95F02}
\definecolor{cSlate}{HTML}{7570B3}
\definecolor{cMagenta}{HTML}{E7298A}

\newcommand{\legline}[2]{\tikz[baseline=-0.5ex]\draw[#1,line width=1.2pt,#2] (0,0) -- (0.7cm,0);}

\begin{document}

\title[Spatially developing drag reduction across a finite oscillating wall]{Spatially developing drag reduction across a finite oscillating wall}

\author*[1]{\fnm{Marco} \sur{Galli}}\email{m.galli-1@tudelft.nl}
\author[2]{\fnm{Nicol\`o} \sur{Scaiola}}\email{n.scaiola@tudelft.nl}
\author[1]{\fnm{Ferdinand F. J.} \sur{Schrijer}}\email{f.f.j.schrijer@tudelft.nl}
\author[1]{\fnm{Fulvio} \sur{Scarano}}\email{f.scarano@tudelft.nl}

\affil*[1]{\orgdiv{Faculty of Aerospace Engineering, Flow Physics and Technology Department}, \orgname{Delft University of Technology}, \orgaddress{\street{Kluyverweg 1}, \postcode{2629 HS}, \city{Delft}, \country{The Netherlands}}}

\affil[2]{\orgdiv{Faculty of Mechanical Engineering, Process and Energy Department}, \orgname{Delft University of Technology}, \orgaddress{\street{Mekelweg 2}, \postcode{2628 CD}, \city{Delft}, \country{The Netherlands}}}

\abstract{The spatial evolution of a turbulent boundary layer at $Re_\tau = 652$ developing across a finite spanwise oscillating wall is investigated with normalised oscillation period of $T^+ \approx 101$ and amplitude $A^+ \approx 7$, yielding a drag reduction of 32\%. The spatial distribution of skin friction is estimated adopting a recently introduced near-wall data nonlinear regression method based on the arctangent function, which circumvents the requirement of high wall-normal resolution to capture the viscous sublayer. The method viability is assessed against reference DNS data, and the trends of measurement errors from experimental data are well reproduced.
The local impact of the control action is examined by evaluating the spatial development of scalar fields of turbulence statistics as well as conditional quadrant analysis to link the onset and recovery transients to turbulence events. The boundary layer response is observed to follow an exponential decay law for the skin friction spatial evolution. The decay length-scale corresponds to $0.6$ boundary layer thicknesses $\delta$ for the onset transient and $0.9\delta$ for the recovery one. The same exponential fit applied to the Reynolds stresses yields a response within $0.5\delta$, with the exception of the shear stress peak, that requires a significantly longer distance.}


\maketitle

\section{Introduction}\label{sec:intro}

Wall shear stress in turbulent boundary layers (TBLs) significantly exceeds its laminar counterpart as a result of wall-normal momentum transport by the turbulent motions of the fluid \citep{orlandi1994}. The reduction of turbulent skin friction still stands as an engineering goal in various industries, due to the compelling need for aerodynamic efficiency and emissions abatement.

A proven active control strategy for a turbulent boundary layer is that of imposing a transverse, oscillatory motion of the wall. First developed by \cite{jung1992}, this technique is generally denoted as `spanwise wall forcing'. It has been explored both in a purely temporal (oscillating wall) or spatial (standing wave) fashion, as well as a combination thereof, leading to the concept of the streamwise-travelling wave \citep{quadrio2009}. The present study focuses on the former, where the entire wall oscillates with spanwise velocity $W_w$ according to a sinusoidal law:
\begin{equation}\label{eq:1}
W_w(t) = A \sin\left(\frac{2\pi}{T}t\right)
\end{equation}
where $A$ and $T$ are the velocity amplitude and period of oscillation, respectively. The normalised skin friction drag reduction $\mathcal{R}$ reads as:
\begin{equation}\label{eq:2}
\mathcal{R} \triangleq \frac{C_{f,0} - C_{f,c}}{C_{f,0}}
\end{equation}
where $C_{f,0}$ and $C_{f,c}$ are the unperturbed and controlled values, respectively. With a suitable choice of the control parameters, $\mathcal{R}$ exceeds 40\%, as reported from numerical simulations \citep{quadrio2004}. Laboratory experiments that have so far been realised yield up to 25\% reduction \citep[see for instance][]{riccowu2004}, mostly due to the challenge in reproducing the conditions of amplitude and period in a physical system.

Control parameters are commonly expressed in non-dimensional form with respect to the friction velocity $u_\tau \triangleq \sqrt{\tau_w / \rho}$ and length-scale $\delta_\nu \triangleq \nu / u_\tau$, with $\tau_w$ indicating the wall shear stress, and $\rho$ and $\nu$ the fluid density and kinematic viscosity, respectively. Therefore, the inner-scaled (hereinafter indicated by the superscript `$+$') amplitude of wall oscillation velocity reads as $A^+ = A / u_\tau$, and the non-dimensional oscillation period as $T^+ = T u_\tau^2 / \nu$. At this stage, it is crucial to remark that in experiments one can only fix the oscillations peak-to-peak displacement $2D_m$, linked to the velocity amplitude $A$ by the relation $A = 2\pi D_m / T$.

Drag reduction has been reported to increase monotonically with $A^+$, while an optimal period for boundary layers is found at $60 < T^+ < 80$ \citep{ricco2004}. The optimum condition reached for a fixed value of $D_m^+$ is therefore lower than the equivalent one at a given value of $A^+$ \citep{riccowu2004}.

\begin{figure}[htbp]
\centering
\begin{minipage}[c]{0.42\textwidth}
\centering
\footnotesize
\setlength{\tabcolsep}{3.5pt}
\begin{tabular}{@{}clccc@{}}
\toprule
& \textbf{Reference} & $T^+$ & $A^+$ & $Re_\theta$ \\
\midrule
\textcolor{cRed}{$\blacksquare$} & \cite{trujillo1997} & 67 & 11.3 & 1500 \\
\textcolor{cBlue}{$\boldsymbol{\times}$} & \cite{choi1998} & \textcolor{blue}{154} & \textcolor{blue}{7.1} & 1190 \\
\textcolor{cGreen}{$\blacklozenge$} & \cite{bogard2000} & 67 & 9.0 & 1400 \\
$\bigcirc$ & \cite{riccowu2004} & 67 & 11.3 & 1400 \\
\legline{cPurple}{solid} & \cite{yudhistira2011} & 100 & 18.0 & 532 \\
\legline{cOrange}{dash dot} & \cite{skote2012} & \multirow{2}{*}{132} & 6.0 & \multirow{3}{*}{503} \\
\legline{cBrown}{dash dot} & & & 12.0 & \\
\legline{cPink}{dash dot} & & 67 & 11.3 & \\
\legline{cTeal}{dotted} & \cite{lardeau2013} & 80 & 12.0 & \textcolor{blue}{1119} \\
\legline{cRust}{dashed} & \cite{skote2019} & \multirow{3}{*}{67} & \multirow{3}{*}{11.3} & 375 \\
\legline{cSlate}{dashed} & & & & 505 \\
\legline{cMagenta}{dashed} & & & & 1400 \\
\botrule
\end{tabular}
\end{minipage}%
\hfill
\begin{minipage}[c]{0.50\textwidth}
\centering
\includegraphics[width=\textwidth]{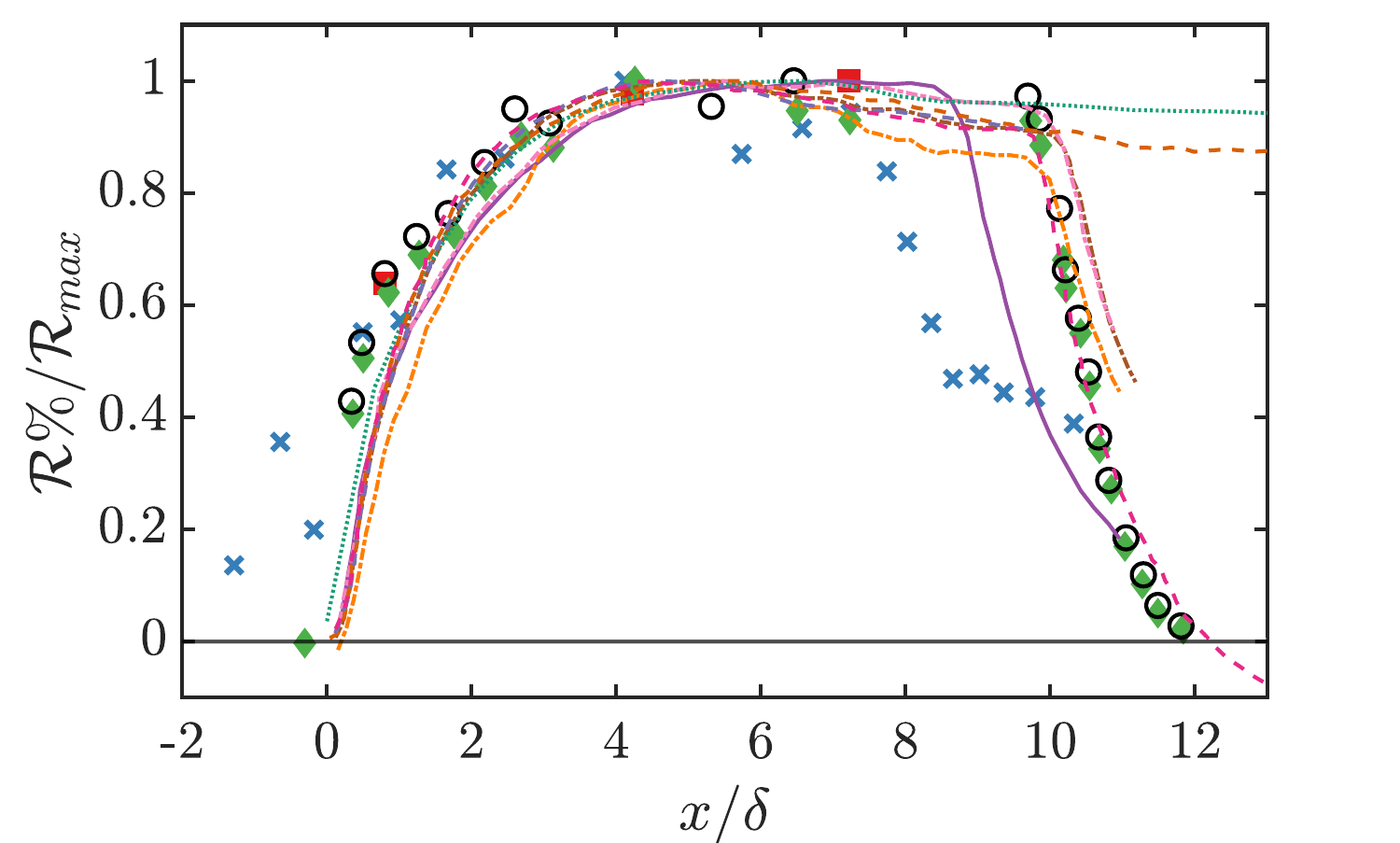}
\end{minipage}
\caption{Normalised drag reduction development across a finite spanwise oscillating wall. Entries in blue are estimated based on the relation $Re_\tau = 1.12 Re_\theta^{0.875}$ \citep{ricco2008}, valid for $Re_\theta < 3000$.}\label{fig:1}
\end{figure}

The above results refer to fully developed conditions obtained along a relatively long segment of a wall. In practical applications, the length of the controlled surface must be finite along the streamwise direction; thus, the boundary layer flow undergoes a spatial transient, both entering and exiting the actuator. Few works have examined such onset and offset, and the pertinent results are summarised in Fig.~\ref{fig:1}. The studies are performed at a Reynolds number based on momentum thickness $\theta$ in the order of $Re_\theta = 10^3$, while oscillation amplitude and period are often selected at the point of maximum drag reduction ($T^+ \approx 100$).

Despite different flow and actuation conditions, the works seem to agree on the onset transient: when normalised between the two reference values (pre-actuation and within actuation), $\mathcal{R}$ reaches asymptotic conditions (viz.\ plateau) after approximately $5$ boundary layer thicknesses $\delta$.

After the cessation of actuation, the data exhibit a more visible scatter of the inception point; furthermore, the asymptotic behaviour is not captured, making it more difficult to infer the transitory length. Given the uncertainties related to the amount of skin friction reduction and the length necessary for reaching an asymptotic value, an approach based on a simplified model is proposed here, which represents the spatial response with a single-exponent decaying law. For the onset transient, the model reads:
\begin{equation}\label{eq:3}
C_f(x) = C_{f,0} - (C_{f,0} - C_{f,c})\left(1 - \mathrm{e}^{-(x - x_0)/\xi_0}\right)
\end{equation}
Here $\xi_0$ plays the role of an ``adaptation length''. The model assumes that at $x = x_0$ (i.e.\ oscillation onset) the wall friction is unperturbed, thus neglecting upstream effects \citep{ricco2021}. As commonly adopted for first-order, linear, time-invariant step response, the transient duration is defined by $\xi$. Introducing the streamwise-varying normalised drag reduction as:
\begin{equation}\label{eq:4}
\mathcal{R}(x) = \frac{C_{f,0} - C_f(x)}{C_{f,0} - C_{f,c}} 
\end{equation}
one can finally express $\mathcal{R}(x)$ as:
\begin{equation}\label{eq:5}
\mathcal{R}(x) = \mathcal{R}\left(1 - \mathrm{e}^{-(x - x_0)/\xi_0}\right)
\end{equation}
From the above, $\mathcal{R}$ reduces its initial value by $1/\mathrm{e}$ at $x = x_0 + \xi_0$, thereby reaching $\approx 63\%$ of the asymptotic value. To provide a more systematic evaluation of the transient, hereafter the related discussion will be cast in said framework. In the same fashion, the recovery transient can be modelled as:
\begin{equation}\label{eq:6}
C_f(x) = C_{f,r} - \Delta C_{f,r}\,\mathrm{e}^{-(x - x_1)/\xi_1}
\end{equation}
where $C_{f,r}$ is the asymptotic value at the end of the recovery transient, $\Delta C_{f,r} = C_{f,r} - C_{f,c}$ and $x_1$ corresponds to the rearmost edge of the actuated segment. Note that, in general, $C_{f,r} \neq C_{f,0}$ and $\xi_0 \neq \xi_1$.

Numerical investigations have dominated the literature on the subject, given the suitability of the DNS technique to model the boundary layer and the oscillating wall \citep[see][for a review]{ricco2021} with no physical limits to amplitude and frequency of actuation. In contrast, the experimental realisation of such flow conditions remains hampered by the challenging conditions imposed by mechanical actuation, severely limited by inertial effects, vibrations, and mechanical friction. Experiments typically feature oscillations of a thin plate achieved via crankshaft systems. \cite{laadhari1994} operated the oscillating wall between 2 and \qty{10}{\hertz} and with a displacement amplitude of \qty{2.5}{\centi\metre}. \cite{choi1998} were able to reach an amplitude of \qty{7}{\centi\metre} for oscillations at \qty{7}{\hertz}. \cite{dicicca2002} conducted measurements in water, hence being able to confine the oscillations to \qty{2.67}{\hertz} at a \qty{2.5}{\centi\metre} amplitude. \cite{kempaiah2020} performed experiments in a wind tunnel where a slab of \qty{40}{\centi\metre} length and width was oscillated at \qty{15}{\hertz} with an amplitude of \qty{1}{\centi\metre}. The present work employs the same architecture and slightly expands the conditions to a frequency of \qty{19}{\hertz} and an amplitude of \qty{2}{\centi\metre}.

Besides mechanical challenges, the condition of a rapidly oscillating surface also poses additional difficulties to the experimental techniques employed to characterise wall friction. \cite{laadhari1994} employed three-dimensional hot-wire probes. \cite{trujillo1997} assessed the near-wall flow behaviour by means of a hot-film probe. They could approach the wall down to a distance of $y^+ = 4$ ($y^+ \triangleq y / \delta_\nu$) due to the physical interference between the actuated wall and the probe. Skin friction was inferred from the linear velocity profile at $y^+ \leq 5$. \cite{choi1998} followed the same approach, gauging the streamwise velocity with a single hot-wire probe. \cite{riccowu2004} utilised a single-component hot-film anemometer, corrected for the bias generated by spanwise oscillations, and a laser Doppler velocimeter in the region $3 \leq y^+ \leq 10$.

Particle Image Velocimetry (PIV) has been proven suitable to provide a single-point or a spatially averaged value of skin friction, fitting near-wall data in the linear range \citep{dicicca2002, kempaiah2020, dacome2024}. However, describing the spatial distribution of skin friction, such as to accurately map the transitory region across a stationary wall and the oscillating segment, highly depends upon the PIV measurement configuration (seeding concentration, optical magnification, particle displacement, etc.), experimental conditions (wall position stability, light reflections mitigation) and data analysis method (cross-correlation or particle tracking).

The best-practice approach for skin friction measurements aims at inferring the velocity at the wall from capturing $4 \div 5$ points of the velocity profile in the linear range. The PIV layout typically requires high-magnification imaging to produce high-resolution images of the tracers near the wall at a distance in the order of \qtyrange{0.1}{1}{\milli\metre}. As a result, the measurement domain is limited to the field of view, requiring in turn that a patchwork of multiple views is needed to cover the wide length of the actuator.

The above has motivated the investigation of methods that permit measurements at lower spatial resolution to fit not only the viscous sublayer but also part of the buffer layer across the near-wall velocity profile, making use of non-linear base functions. The topic has been covered recently by \cite{zhou2024}, who attempted to determine the wall friction based on an analytical wall-law. They show that including information beyond the viscous sublayer can significantly improve the accuracy of $\tau_w$, both in DNS and experiments. The formulation is based on the hypothesis of a canonical zero-pressure-gradient turbulent boundary layer and requires the evaluation or estimation of a considerable number of parameters.

Differently, \cite{hendriksen2025} have proposed an empirical non-linear function to reduce the spatial resolution requirements for skin friction measurements when employing large-scale 3D particle tracking. In their study they deal with helium-filled soap bubbles of sub-millimetre diameter, rendering the direct measurement of the linear layer impossible in most aerodynamic flows.

The present work employs planar PIV with a specific arrangement that slides streamwise the field of view to include the wide spatial region of the oscillating actuator. It must be reminded that an inherent limitation of planar PIV is the absence of information concerning the spanwise component of the wall shear stress. This contribution is relevant in wall-bounded flows subjected to transverse forcing; however, it does not impact considerations on drag reduction nor statistics, since it possesses zero mean. Furthermore, its behaviour can be recovered analytically, since it presents an excellent match with the laminar solution to Stokes' second problem, at least for $T^+$ close to the optimum.

The data analysis follows and expands the approach of \cite{hendriksen2025}, aiming at a quantitative characterisation of the skin friction coefficient and its spatial variation across the oscillating wall. The first part of the study is devoted to characterising the accuracy, robustness and generality of such non-linear function for the canonical case of a turbulent boundary layer over a flat plate.

Further, the experiments describe the spatially varying properties of the TBL when the flow transits across the edges of a finite oscillating plate. The spatial distribution of the skin friction coefficient along the streamwise direction $C_f(x)$ is aimed at, alongside the relevant properties of the turbulent boundary layer, notably the Reynolds stresses. Finally, a quadrant analysis links velocity fluctuations to relevant motions within the boundary layer.

\section{Methodology}\label{sec:methodology}

\subsection{Experimental setup}\label{subsec:setup}

Experiments were performed in a low-speed open-return wind tunnel (W-tunnel) at the aerodynamics laboratories of Delft University of Technology. The facility features a 9:1 contraction ratio with a test section of $\qty{0.4}{\metre} \times \qty{0.4}{\metre}$ into a square channel of \qty{1.4}{\metre} length. The tunnel is operated at \qty{3.5}{\metre\per\second}, with a freestream turbulence intensity below 1\%.

\begin{figure}[htbp]
\centering
\includegraphics[width=\textwidth]{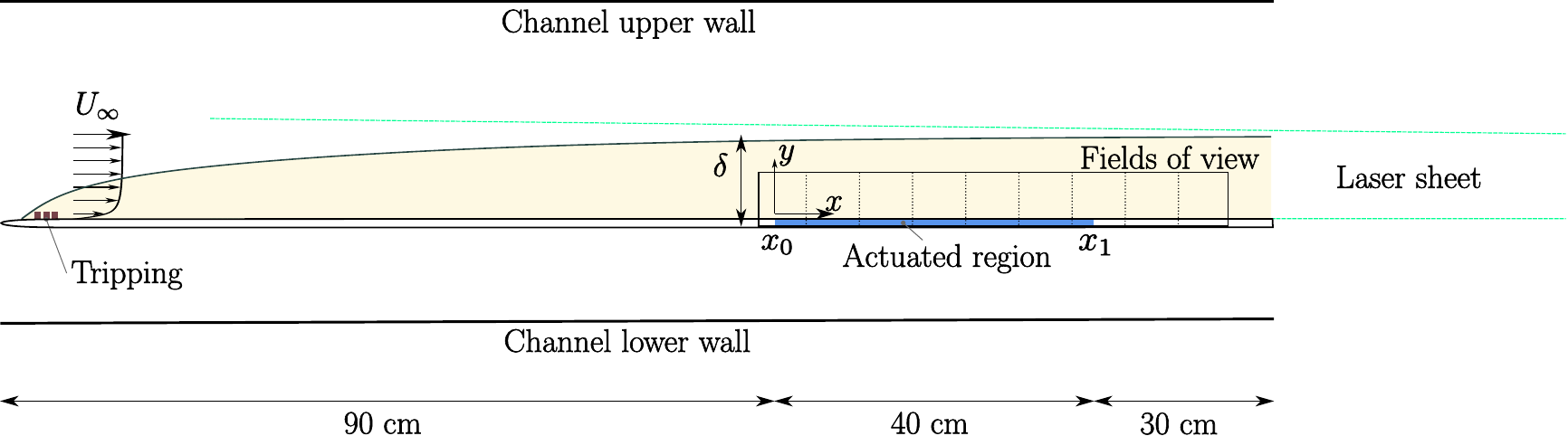}
\caption{Schematic of the experimental setup.}\label{fig:2}
\end{figure}

A flat plate with an elliptical leading edge is installed \qty{0.1}{\metre} above the bottom wall. A laminar boundary layer undergoes forced transition by a tripping device comprising distributed \qty{1}{\centi\metre} cubic elements spaced by \qty{3}{\centi\metre} each, following previous works from \cite{limapereira2020} and \cite{kempaiah2020}. The resulting turbulent boundary layer develops for a length of approximately \qty{1}{\metre} with negligible pressure gradient, until encountering a \qty{40}{\centi\metre}-long oscillating segment followed by another stationary part \qty{0.3}{\metre} long.

The oscillating plate is a \qty{5}{\milli\metre}-thick composite sandwich structure with two thin layers of carbon fibre enclosing a styrene foam. Its width is \qty{60}{\centi\metre}, and it weighs \qty{50}{\gram}. The plate is connected to a slider-crank mechanism powered by a brushless DC Maxon motor operated with a MIP100 positioning controller. The drive system features a counterweight to minimise vibrations.

\begin{figure}[htbp]
\centering
\includegraphics[width=0.75\textwidth]{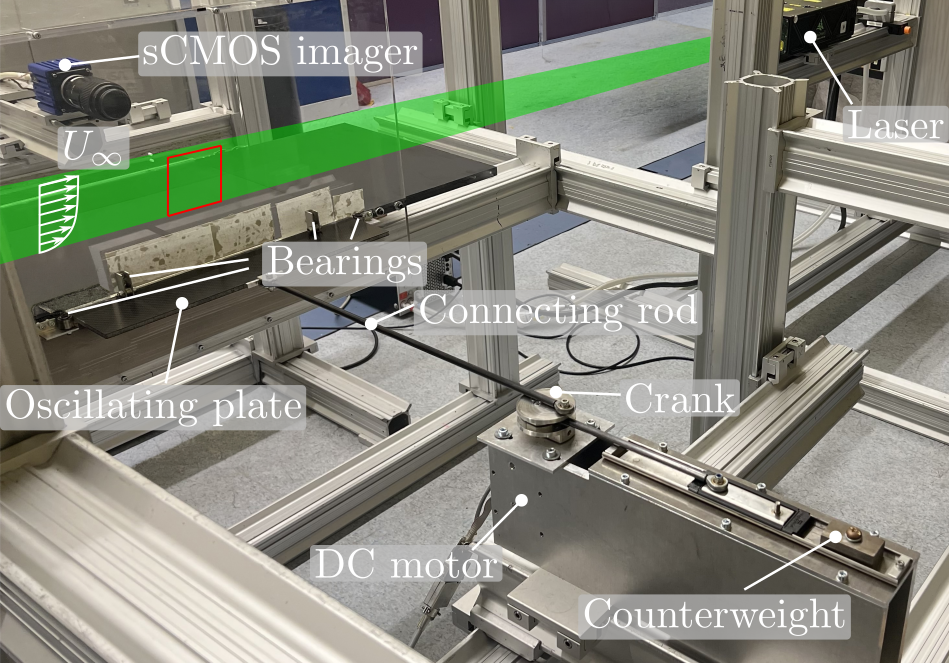}
\caption{Annotated photograph of the experimental apparatus.}\label{fig:3}
\end{figure}

The apparatus produces sinusoidal oscillations up to a frequency of \qty{19}{\hertz}. The peak-to-peak amplitude can be varied by changing the radial position of the rod-crank joint, currently set to \qty{5}{\centi\metre}, yielding a maximum velocity of \qty{2.4}{\metre\per\second}. Under these operating conditions the electrical power input to the motor is approximately \qty{120}{\watt}.

The interface gaps and steps are kept to a minimum of approximately \qty{0.2}{\milli\metre} owing to the apparatus assembly tolerances. A sketch of the experimental setup is provided in Fig.~\ref{fig:2}, whereas a photograph of the experimental apparatus is displayed in Fig.~\ref{fig:3}. A video recording of the actuated surface is made available as supplementary material (see the dedicated section on data availability for more information).

\subsection{PIV measurements}\label{subsec:piv}

Planar PIV is used to obtain the velocity components in the streamwise--wall-normal ($x$--$y$) plane. Fog droplets with \qty{1}{\micro\metre} diameter, produced from a water--glycol mixture by a SAFEX Fog 2010+ smoke generator, are employed as seeding particles. To obtain ideal conditions of homogeneous concentration \citep{kahler2004} for PIV measurements, the seeded flow is recirculated until the desired seeding density is reached.

Illumination is delivered by means of a Quantel Nd:YAG Evergreen 200 laser ($2 \times \qty{200}{\milli\joule}$/pulse, $\lambda = \qty{532}{\nano\metre}$). A \qty{1.5}{\milli\metre}-thick and \qty{70}{\milli\metre}-wide laser sheet is obtained by means of a combination of cylindrical and spherical lenses. The laser light is carefully aligned with the wall to minimise the amount of light impingement causing spurious reflections. The light scattered by the particle tracers is captured with a LaVision Imager sCMOS camera ($2560 \times 2160~\mathrm{px}$) equipped with a Nikkor AF-S $f = \qty{105}{\milli\metre}$ lens at an aperture of $f_\# = 8.0$. At a working distance of \qty{1}{\metre}, the optical magnification of $0.19$ yields a field of view (FOV) of $87 \times 74~\mathrm{mm}^2$. Given the slender aspect ratio of the domain of interest, the FOV needs to be repeated several times by translating the camera along the $x$-axis on a rail, covering a total length of \qty{75}{\centi\metre} with limited sacrifice to the spatial resolution ($\approx 30~\mathrm{px/mm}$) to infer the wall friction distribution.

Digital synchronisation between laser pulses and image acquisition is ensured by a LaVision programmable timing unit (PTU), controlled through DaVis 10.2. Table~\ref{tab:1} summarises the hardware employed. The pulse delay is set to $\Delta t = \qty{150}{\micro\second}$, to yield a freestream particle image displacement of approximately \qty{0.53}{\milli\metre} ($\approx 16~\mathrm{px}$).

\begin{figure}[htbp]
\centering
\includegraphics[width=\textwidth]{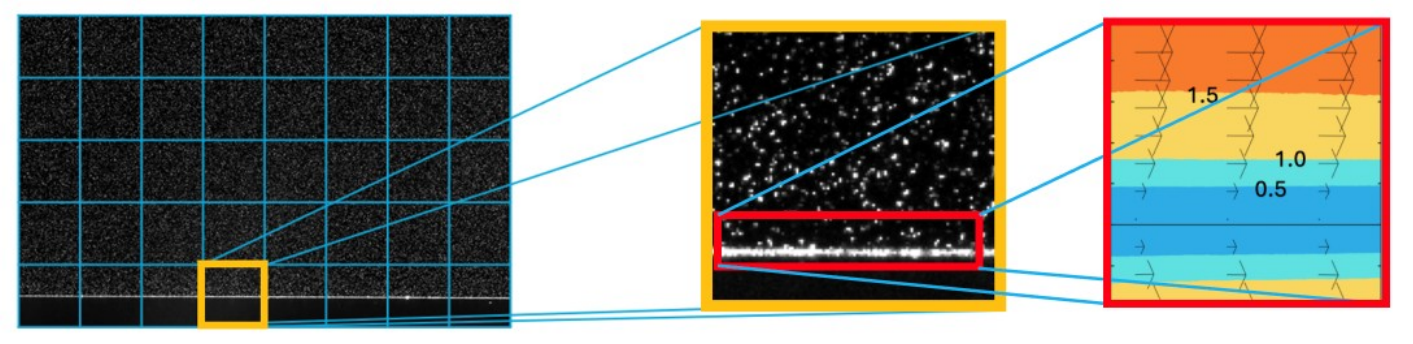}
\caption{Particle image recording from a single FOV (left) and a detail of it near the wall (middle). Time-averaged velocity vectors (right) with window size of $32 \times 8~\mathrm{px}$ (height of \qty{0.3}{\milli\metre}, $3\delta_\nu$). Labels indicate the velocity at colour contours. The vertical axis is stretched by a factor 10 for visibility.}\label{fig:4}
\end{figure}

At each FOV, a set of 1000 double-frame recordings is acquired. An example is shown in Fig.~\ref{fig:4}, alongside a detail of the wall reflections. Before interrogation, image pre-processing involves normalisation with average intensity and subtracting the minimum pixel intensity over time from a sequence of 7 frames.

The particle image motion is analysed with multi-pass cross-correlation including window deformation and grid refinement \citep{scarano2000}. For the turbulence statistics, the final window is Gaussian-weighted with $24 \times 24~\mathrm{px}$ ($9 \times 9\,\delta_\nu$). An overlap factor of 75\% yields a vector pitch of approximately \qty{0.2}{\milli\metre} ($2\delta_\nu$). For the measurement of wall friction a smaller and non-isotropic ($32 \times 8~\mathrm{px}$, $12 \times 3\,\delta_\nu$) window \citep{scarano2003} has been adopted, with an overlap of 50\% along the wall-normal direction, yielding a vector pitch of $4~\mathrm{px}$ ($1.5\delta_\nu$).

The rightmost plot in Fig.~\ref{fig:4} also illustrates the no-slip condition at the wall not being achieved precisely due to the spatial averaging effect of the cross-correlation operator.

\begin{table}[htbp]
\caption{PIV system hardware details}\label{tab:1}%
\begin{tabular}{@{}lcr@{}}
\toprule
\multirow{2}{*}{\textbf{Seeding}} & Particles type & Fog \\
 & Particle diameter & \qty{1}{\micro\metre} \\
\midrule
\multirow{4}{*}{\textbf{Illumination}} & Technology & Nd:YAG laser \\
 & Energy & $2 \times \qty{200}{\milli\joule}$ \\
 & Pulse repetition rate & \qty{15}{\hertz} \\
 & Mode of operation & Q-switched \\
\midrule
\multirow{8}{*}{\textbf{Recording}} & Sensor type & sCMOS \\
 & Sensor size & $2560 \times 2160~\mathrm{px}^2$ \\
 & Pixel pitch & \qty{6.5}{\micro\metre} \\
 & Digital output & 16 bit \\
 & Method & Double frame \\
 & Lens and aperture & \qty{105}{\milli\metre}, $f_\# = 8.0$ \\
 & Pulse delay & \qty{150}{\micro\second} \\
 & Magnification factor & 0.19 \\
\botrule
\end{tabular}
\end{table}

Image recording and interrogation parameters are summarised in Table~\ref{tab:2}. After cross-correlation, each instantaneous velocity vector field is inspected using the universal outlier detector \citep{westerweel2005}. The average detection rate is below 1\% in the flow above the wall and approximately 5\% in the near-wall region ($y^+ \leq 10$, $y \leq \qty{0.9}{\milli\metre}$).

\begin{table}[htbp]
\caption{Image acquisition and processing details}\label{tab:2}%
\begin{tabular}{@{}lcr@{}}
\toprule
\multirow{4}{*}{\textbf{Acquisition}} & Number of images & 2000 (at each location) \\
 & Image resolution & 30 pixels/mm \\
 & Seeding density & 0.035 ppp \\
 & Freestream particle displacement & \qty{0.53}{\milli\metre} (16 px) \\
\midrule
\multirow{7}{*}{\textbf{Processing}} & \multirow{2}{*}{Pre-processing} & Normalise by average intensity \\
 & & Subtract minimum pixel intensity \\
 & Interrogation window & $24 \times 24~\mathrm{px}^2$ / $32 \times 8~\mathrm{px}^2$ \\
 & Overlap & 75\% / 50\% \\
 & \multirow{2}{*}{Vector pitch} & \qty{0.2}{\milli\metre} ($\approx 5~\mathrm{vec/mm}$, $2.21\delta_\nu$) \\
 & & \qty{0.13}{\milli\metre} ($\approx 7~\mathrm{vec/mm}$, $1.5\delta_\nu$) \\
 & Dynamic spatial range & 433 / 667 \\
\botrule
\end{tabular}
\end{table}

\subsection{Wall shear stress estimation}\label{subsec:wss}

The measurement of the wall shear stress $\tau_w$ requires the estimation of the velocity gradient at the wall. Several factors hamper the measurement of such quantity in the close proximity of the wall \citep{kahler2012}, namely: optical disturbances (mostly reflections, see Fig.~\ref{fig:4}), a limited spatial resolution (in turn related to particle image size and density) and measurement noise (cross-correlation in highly sheared conditions \citep{scarano2002}). Such limitations are often circumvented by extrapolating the velocity gradient at the wall from a thin region slightly above it. The best-practice approach for skin friction evaluation from PIV data is that of fitting linearly $4 \div 5$ datapoints in the viscous sublayer as reported by \cite{kempaiah2020} among others. Adopting particle tracking in place of cross-correlation analysis offers an increase in spatial resolution at the expense, however, of a lower velocity dynamic range \citep{dacome2024}. In the latter study, however, a significantly higher optical magnification ($M = 0.74$) was applied, covering an exceedingly small field of view, in contrast with the need to measure along an extended domain (\qty{75}{\centi\metre}, or $\approx 13\delta$ of the present study). A pivotal uncertainty source is constituted by the wall location \citep{kahler2012}, which can severely affect the estimation of the skin friction. For a mechanically oscillating wall, system vibrations, even when minimal, affect wall position estimation. As a result, a data gap near the wall, visible in Fig.~\ref{fig:4}, needs to be accounted for.

Due to the above, the linear fit is only viable when the gap does not erode excessively the viscous sublayer (typically $2 \leq y^+ \leq 5$), corresponding to \qtyrange{0.18}{0.45}{\milli\metre} in the present case. The limited number of datapoints in the linear range and their fidelity result in a limited accuracy of the skin friction estimation by a linear fit. However, as shown in the remainder, expanding the data range of the linear fit beyond the viscous sublayer rapidly increases the bias error, which underestimates the gradient at the wall (see Fig.~\ref{fig:9}), due to the velocity profile significantly deviating from the theoretical relation $u^+ = y^+$ beyond $y^+ = 5$.

\subsubsection{Non-linear modelling}\label{subsubsec:nonlinear}

A non-linear fit based on the arctangent function has been recently investigated by \cite{hendriksen2025} in the framework of skin friction estimation for large-scale three-dimensional measurements around stationary solid objects. In that context, 3D measurements performed with sub-millimetre helium-filled soap bubbles \citep{faleiros2019} prevent the assessment of the first millimetre in the vicinity of the solid object, therefore yielding a data gap even more pronounced than in the present case. The mean streamwise velocity $\bar{u}(y)$ is therefore modelled as an arctangent function along the wall-normal coordinate. Hereinafter, the operator $\overline{\cdots}$ identifies the mean component of velocity:
\begin{equation}\label{eq:7}
\bar{u}(y) \approx a \arctan\left(b(y + c)\right)
\end{equation}
Such fitting model encompasses three parameters with a clear physical meaning: $a$ scales the magnitude of the freestream velocity; $b$ relates to the steepness of the velocity profile at the wall, thus its gradient; finally, $c$ represents the wall position. Provided that a least-squares optimisation can be carried out, fitting the arctangent to the measured data, the velocity gradient at the wall $\left.\partial \bar{u}/\partial y\right|_{y=y_{\mathrm{wall}}}$ can be recovered analytically as:
\begin{equation}\label{eq:8}
\left.\frac{\partial \bar{u}}{\partial y}\right|_{y=y_{\mathrm{wall}}} = \left.\frac{a\,b}{1 + b^2(y + c)^2}\right|_{y=y_{\mathrm{wall}}} = a\,b
\end{equation}
The skin friction coefficient follows from the above as:
\begin{equation}\label{eq:9}
C_f = \frac{\mu \left.\partial \bar{u}/\partial y\right|_{y=y_{\mathrm{wall}}}}{\frac{1}{2}\rho U_\infty^2} = \frac{\nu}{2}\frac{a\,b}{U_\infty^2} 
\end{equation}
where $\mu$ is the fluid dynamic viscosity. As discussed in \cite{hendriksen2025}, the nonlinear character of the arctangent function matches many features of the boundary layer velocity profile, namely:
\begin{itemize}
\item The velocity is zero at the origin (i.e.\ the wall), resulting in the no-slip condition ($\lim\limits_{y \to y_{\mathrm{wall}}} \bar{u}(y) = 0$) being automatically satisfied.
\item The gradient becomes constant at the wall, mimicking the linear behaviour in the viscous sublayer ($\left.\frac{\partial \bar{u}}{\partial y}\right|_{y=y_{\mathrm{wall}}} = \text{const}$).
\item The velocity increases monotonically moving away from the wall ($\frac{\partial \bar{u}}{\partial y} > 0$).
\item The gradient vanishes far from the wall, and the velocity tends to an asymptotic value ($\lim\limits_{y \to \infty} \frac{\partial \bar{u}}{\partial y} = 0$).
\end{itemize}

Given the above, it is expected that the arctangent function can encompass a larger wall-normal range by including the nonlinear region. As a result, more datapoints participate in the estimation with benefits in terms of precision and accuracy of wall gradient estimations. It should be retained in mind, however, that such approach is not intended to model the velocity distribution across the entire boundary layer, but only to reduce the truncation errors when extending the data range beyond the linear sublayer.

For a full description of the boundary layer, a well-known approach aiming at fitting the data using an analytical composite profile has been proposed \citep{chauhan2009}. However, also the latter requires the input of several fitting parameters in the connection of analytical functions and is usually intended for a fully developed and uncontrolled boundary layer.

\subsubsection{Numerical assessment}\label{subsubsec:numerical}

The use of the arctangent fit is scrutinised here with the purpose of determining its limits of validity and providing an estimation of the error. Data from a direct numerical simulation (DNS) of a turbulent boundary layer \citep{schlatter2010} at $Re_\theta = 670$ is taken as reference. The output of a PIV measurement is simulated considering a window size of $4\delta_\nu$ or \qty{0.36}{\milli\metre}. The main properties of the boundary layer are illustrated in Table~\ref{tab:3}.

\begin{table}[htbp]
\caption{Boundary layer properties from the DNS database}\label{tab:3}%
\centering
\begin{tabular}{@{}ccccc@{}}
\toprule
$C_f$ & $\delta$ & $Re_\theta$ & $u_\tau$ & $\delta_\nu$ \\
 & [\unit{\milli\metre}] & & [\unit{\metre\per\second}] & [\unit{\milli\metre}] \\
\midrule
0.0048 & 26 & 670 & 0.15 & 0.10 \\
\botrule
\end{tabular}
\end{table}

\begin{figure}[htpb]
    \centering
    \includegraphics[width=0.6\textwidth]{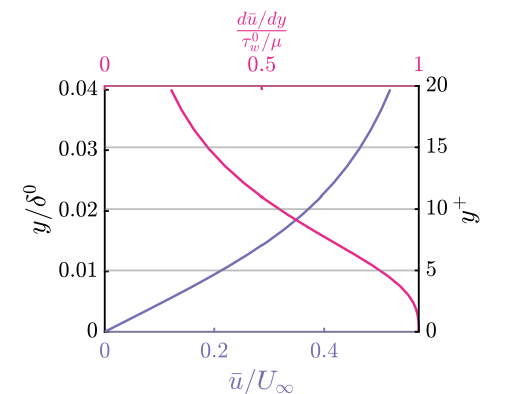}
    \caption{Normalised velocity (purple, bottom axis) and wall-normal gradient (fuchsia, top axis) represented in outer- and inner-scaled (right axis) coordinates.}\label{fig:5}
\end{figure}

The velocity profile $\bar{u}(y)$ as well as the wall-normal gradient are depicted in Fig.~\ref{fig:5}. Although the velocity profile appears well approximated by a linear function in the layer from the wall up to five viscous units, the wall-normal derivative indicates that at such height the value at the wall is already underestimated by 10\%. Therefore, a linear estimator is likely to return a value underestimated by 5\% due to spatial averaging.

The effect of a velocity measurement near the wall is simulated by mirroring the velocity profile below the wall, because of light rays' reflection. The cross-correlation analysis can be modelled as a sliding average top-hat filter, of width corresponding to the window height. Random errors are not introduced yet, focusing the attention initially on the bias error.

\begin{figure}[htbp]
\centering
\includegraphics[width=\textwidth]{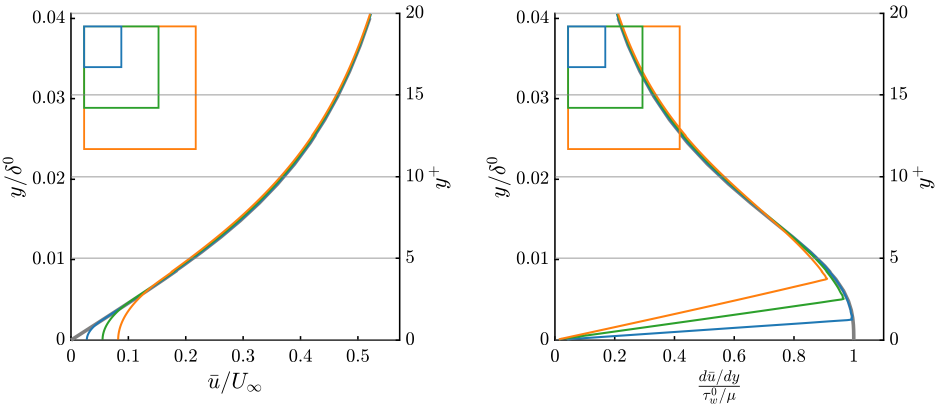}
\caption{Effect of finite spatial resolution on velocity profile (left) and wall-normal gradient (right). Interrogation windows of 2.5 (blue), 5.0 (green) and 7.5 (yellow) $\delta_\nu$ are depicted.}\label{fig:6}
\end{figure}

Figure~\ref{fig:6} illustrates such effects on the velocity and wall gradient distribution. For a window size of $5\delta_\nu$ (\qty{0.45}{\milli\metre}) the velocity profile is affected already at approximately $2.5\delta_\nu$. Also, the maximum level of the gradient is reduced by 5\%.

The linear and arctangent fits are applied in a data range bounded on the lower side (lower bound $y_l^+$) by the window size and in principle unbounded on the upper side. Yet an upper bound ($y_u^+$) needs to be chosen that delimits the range of data used for the fit, yielding the fitting range $[y_l^+, y_u^+]$.

The fidelity of a functional fit in the near-wall region clearly depends upon the available data range and selected fitting range ($y^+ \in [y_l^+, y_u^+]$) and the choice of the function. For the sake of generality, no hypothesis is made here on the position of the wall. In turn, the zero velocity at the wall (no-slip condition) is not imposed with the present approach.

\begin{figure}[htbp]
\centering
\includegraphics[width=\textwidth]{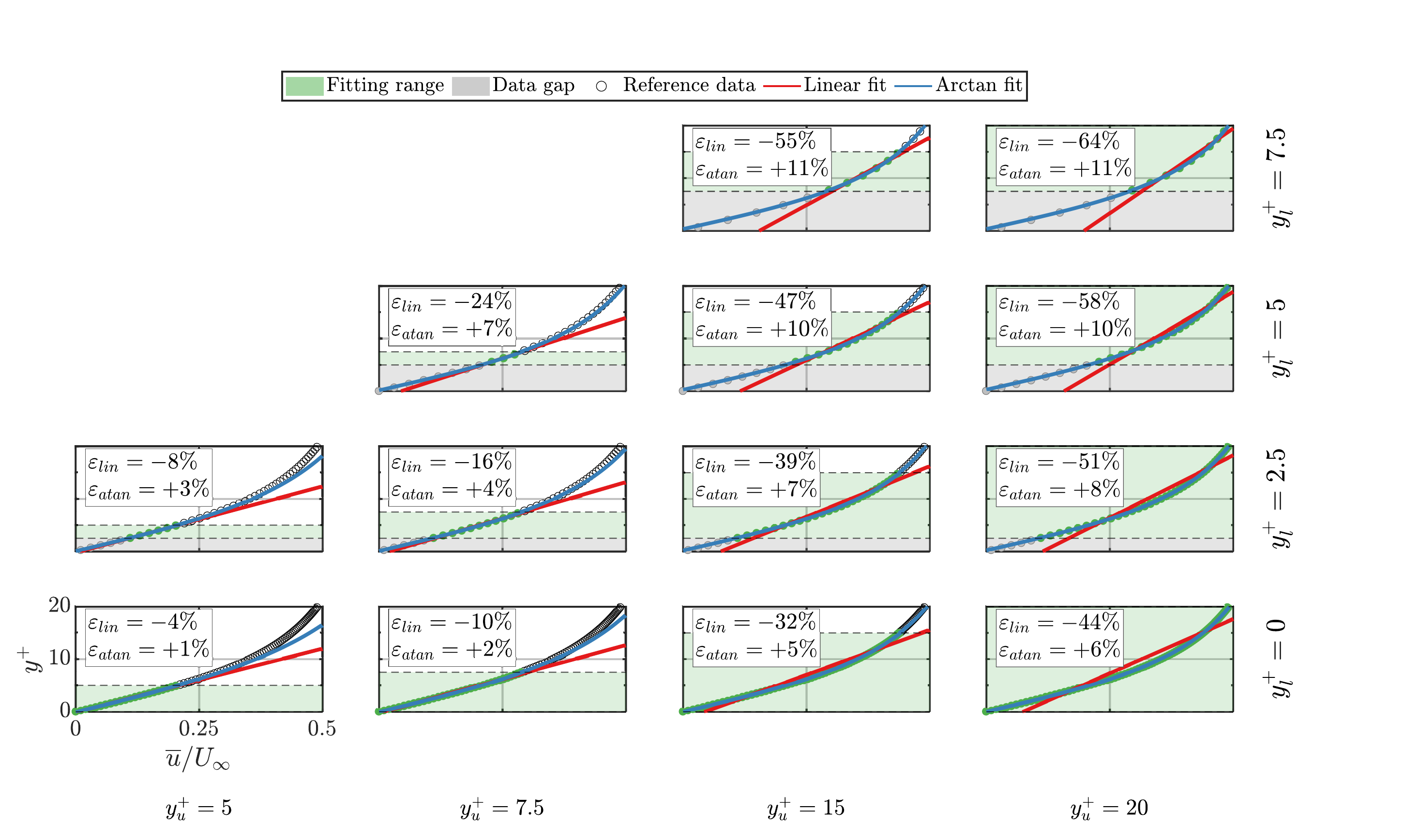}
\caption{Velocity profile obtained with linear (red line) and arctangent fit (blue line) of reference data from DNS (black circles). The gap height is varied from the bottom to top row. The fitting range (green band and labels) upper bound increases from the left to the right column. The legend reports the relative error $\varepsilon$ (percent) of the skin friction estimation from the fit.}\label{fig:7}
\end{figure}

Following \cite{hendriksen2025}, the upper and lower bounds are varied parametrically and the resulting error of $C_f$ is reported. The trends, displayed in Fig.~\ref{fig:7}, elucidate the effect of measurement resolution on the skin friction estimation. The wall gap height is increased from the bottom to the top row, simulating conditions of increasingly coarser resolution. The fitting range upper bound is increased accordingly, and the overall extent of the fitting range is also considered. The latter increases from the left to the right column.

Conditions of a fully resolved velocity field are represented in the bottom-left corner, for a vanishing gap and minimum fitting range ($[y_l^+, y_u^+] = [0, 5]$). Here the skin friction is estimated with the lowest error with the linear fit (4\%) as well as with the arctangent fit (1\%). Increasing the height of the fitting range (moving from left to right along the bottom row), the linear fit becomes rapidly inaccurate, yielding an error in excess of 40\% when $y_u^+ = 20$. This is largely expected, since $\bar{u}(y)$ significantly departs from the linear behaviour beyond $y^+ = 5$. The error committed with the arctangent fit also grows, but comparatively less, and it remains below 6\% at the edge of the considered fitting range. Interestingly, the bias error has opposite sign for the two functions, with the linear fit systematically underestimating the skin friction, whereas the arctangent overestimates it. Finally, increasing the gap height (resulting from measurements at coarser resolution) produces a more rapid increase of the error from both estimators. As a reference, for the case ($[y_l^+, y_u^+] = [5, 15]$), errors of 47\% and 10\% are returned by the linear and arctangent fits, respectively.

\begin{figure}[htbp]
\centering
\includegraphics[width=0.95\textwidth]{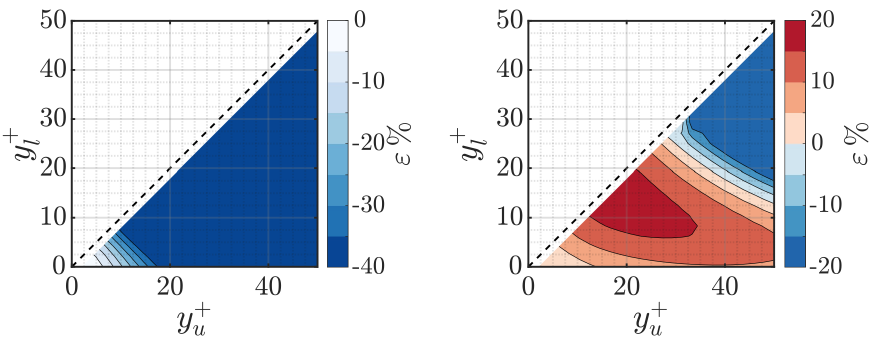}\\[1ex]
\includegraphics[width=0.95\textwidth]{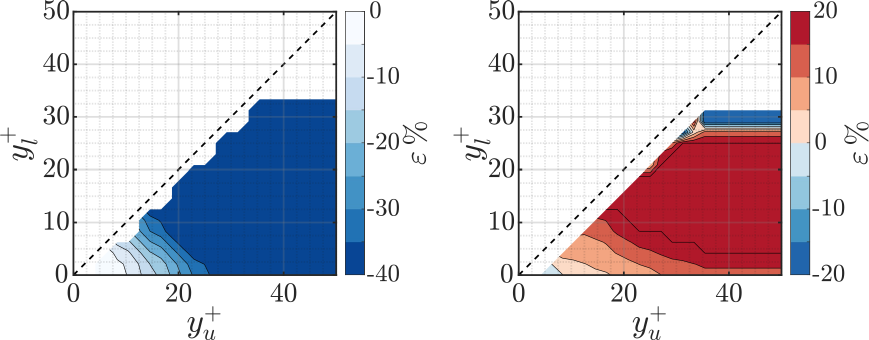}
\caption{Skin friction error as a function of upper and lower bounds of linear (left) and arctangent (right) fit. The upper row refers to the DNS dataset. The lower row reports the same maps estimated from the PIV measurements.}\label{fig:8}
\end{figure}

A summarised view of the error dependence upon the gap height and fitting range is displayed as a contour map in Fig.~\ref{fig:8}. Isolines of the error for the linear fit (left) are inclined at approximately \qty{-45}{\degree}, indicating that the error grows the fastest when the lower and upper bounds are increased equally, i.e.\ when the fitting range is shifted upwards. Overall, the error exceeds 40\% for a gap height of $10\delta_\nu$. The error map for the arctangent fit (right) shows a non-monotonic behaviour, with the error increasing to a maximum of 20\% until the point $[y_l^+, y_u^+] = [10, 20]$. Beyond such point the error declines and crosses the zero value at approximately $[y_l^+, y_u^+] = [20, 30]$, eventually becoming negative further on. A favourable feature is that the error isolines behave quasi-horizontally in the lower part of the diagram ($y_l^+ < 10$), with the error only growing mildly when the upper bound is increased. For verification, a similar exercise is performed with the data obtained from the current experiments. The general trends and values compare favourably, although some differences can be appreciated. In particular, considering the realistic conditions of $[y_l^+, y_u^+] = [5, 15]$, error levels of approximately 30\% and 9\% are observed for the linear and arctangent fits, respectively.

\subsubsection{De-biasing the skin friction estimate}\label{subsubsec:debiasing}

\begin{figure}[htbp]
\centering
\includegraphics[width=\textwidth]{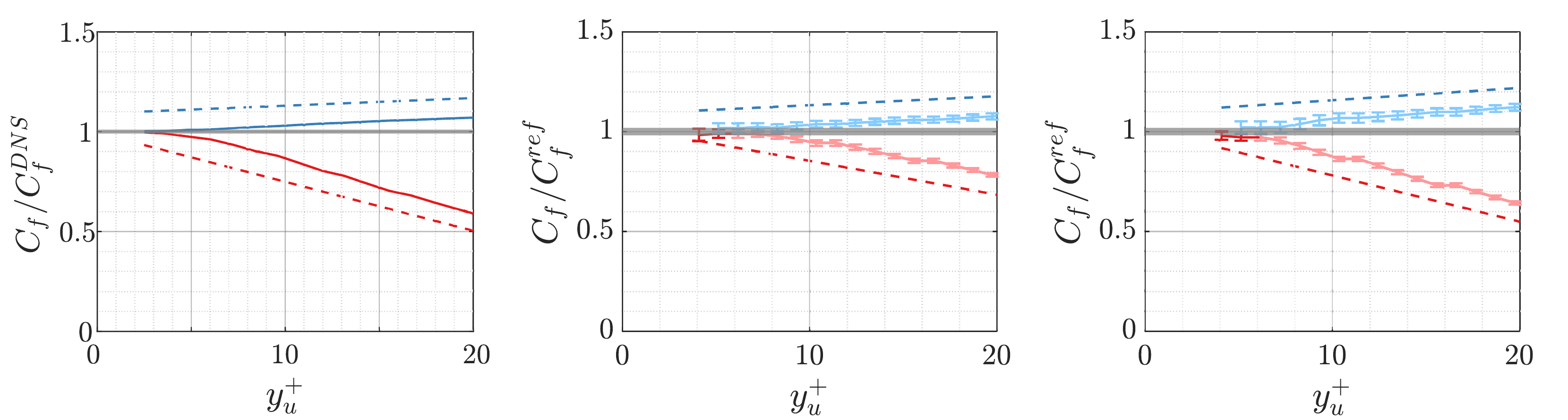}
\caption{Estimated skin friction coefficient from linear fit (red solid) and arctangent fit (blue solid) varying the upper bound of the fitting range $y_u^+$. The thick grey line represents the ground truth from DNS data (left). Data from experiments for $x \in [240, 315]~\mathrm{mm}$ over a stationary (middle) and oscillating (right) wall. Here the thick grey line is only indicative of the expected value of $C_f$. Error bars represent the standard deviation from measurements obtained at varying streamwise positions.}\label{fig:9}
\end{figure}

As previously discussed and evident from Fig.~\ref{fig:8}, the trends of $C_f$ vs.\ $y_u^+$ exhibit an error that increases monotonically within the range up to $y_u^+ = 20$. Furthermore, the error of the linear and arctangent fits displays opposite behaviours. Finally, the error appears to vanish in the limit of $y_u^+ \to 0$. It can be hypothesised that the two error trends intersect in close proximity to the exact, unbiased value towards such limit. This condition is illustrated in Fig.~\ref{fig:9} (left), where the blue solid line is the error committed with the arctangent fit for a varying value of the upper bound (the lower is kept fixed at $2.5\delta_\nu$). The corresponding error with the linear fit is given by the red line. A linear fit of both functions systematically yields an intersection at a rather constant value of $C_f$, departing minimally from the DNS reference value. Verification of this condition is shown in Fig.~\ref{fig:9} (middle) and Fig.~\ref{fig:9} (right), where the same procedure is applied to the data obtained experimentally in the conditions of a flat-plate turbulent boundary layer over a stationary or spanwise oscillating wall, respectively. Although the absolute value of the errors and their slope slightly differ from the numerical example, the conditions of quasi-linearity and stability of their intersection seem to be respected.

From Fig.~\ref{fig:9} it can be inferred that in the range $[0, 20]\delta_\nu$, where the velocity profile already departs largely from the linear behaviour, the bias error increases approximately linearly with the linear fit at a rate of $-10^{-4} C_f / \delta_\nu$ (dashed red line), and approximately five times less ($2 \times 10^{-5} C_f / \delta_\nu$) for the arctangent fit (dashed blue line). As a result, for the latter, the error may remain bounded to within 5\% when the data range extends up to twenty viscous units. This condition marks a significant advantage of using a nonlinear fit when performing PIV experiments with moderate wall-normal resolution.

\begin{figure}[htbp]
\centering
\includegraphics[width=0.9\textwidth]{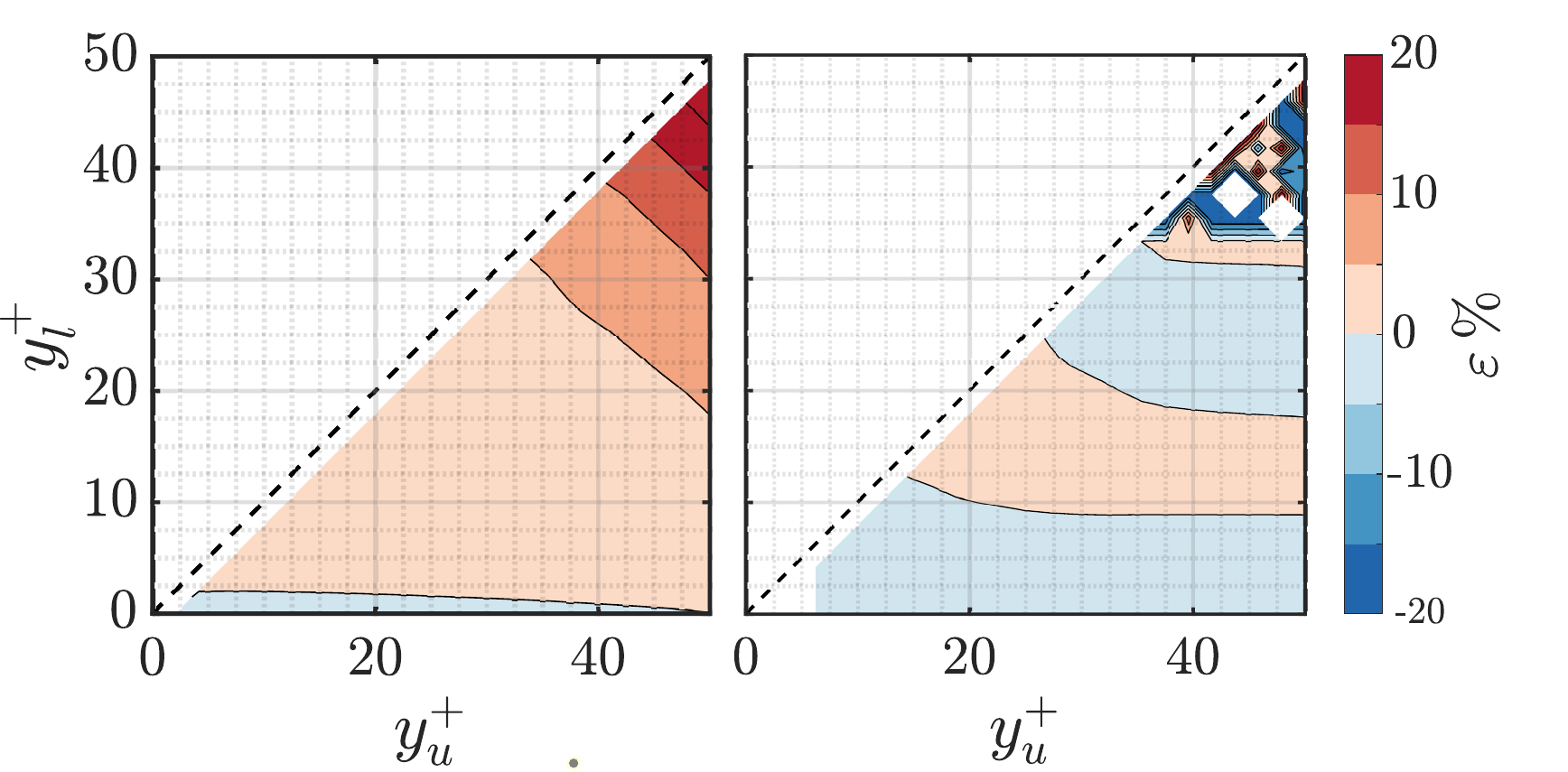}
\caption{Relative $C_f$ error map from the de-biased estimate produced by extrapolated intersection of linear and arctangent fit errors. Left: based on DNS data. Right: based on current measurements.}\label{fig:10}
\end{figure}

The error map as a function of $y_l^+$ and $y_u^+$ is calculated again using the estimate of $C_f$ from the intersection of linear and arctangent estimates of skin friction; the result is reproduced in Fig.~\ref{fig:10}.

When examining the results using the DNS data (Fig.~\ref{fig:10}, left), a wide plateau region bounded within 5\% is revealed. Sampling the value at $[y_l^+, y_u^+] = [5, 15]$ yields an error below 2\%, which can be considered rather accurate in the context of laboratory experiments. When reproducing the same map from the data in the actual experiments (Fig.~\ref{fig:10}, right), the pattern is slightly different, yet a wide region with error bounded by $\pm 5\%$ is observed, which preludes to measurements of skin friction with a lower bias error even when using a wider fitting range across the velocity profile.

It should be retained in mind that the above discussion, however, does not include the effect of random errors as well as other forms of bias errors associated with PIV data analysis, such as peak locking \citep{westerweel1997}. Therefore, a conclusive statement on the subject should also include these sources of error along with a full account of the measurement uncertainty in highly sheared conditions.

\section{Results}\label{sec:results}

\subsection{Incoming boundary layer}\label{subsec:incoming}

The turbulent boundary layer is characterised at $x = \qty{0}{\milli\metre}$, just upstream of the oscillation onset, in terms of first- and second-order statistics. A collection of properties is summarised in Table~\ref{tab:4}.

\begin{table}[htbp]
\caption{Reference boundary layer properties at $x = \qty{0}{\milli\metre}$}\label{tab:4}%
\begin{tabular}{@{}ccccccccc@{}}
\toprule
$U_\infty$ & $\delta$ & $\delta^*$ & $\theta$ & $H$ & $Re_\theta$ & $u_{\tau,0}$ & $Re_\tau$ & $\delta_\nu$ \\
{[\unit{\metre\per\second}]} & {[\unit{\milli\metre}]} & {[\unit{\milli\metre}]} & {[\unit{\milli\metre}]} & & & {[\unit{\metre\per\second}]} & & {[\unit{\milli\metre}]} \\
\midrule
3.5 & 59 & 6.2 & 4.3 & 1.43 & 994 & 0.17 & 652 & 0.09 \\
\botrule
\end{tabular}
\end{table}

They have been inferred through a composite fit procedure on the mean streamwise velocity \citep{chauhan2009}. Hereinafter, $\delta^*$ indicates the displacement thickness, $H \equiv \delta^* / \theta$ the shape factor and $Re_\tau \equiv u_\tau \delta / \nu$ the friction Reynolds number.

The friction coefficient of the baseline flow obtained by fitting the velocity profile is $C_{f,0} = 0.0047$; the arctangent-based estimate yields an agreement within 0.1\%. Employing the empirical Coles--Fernholz relation $C_f = 2\left[1/\kappa\,\ln(Re_\theta) + C\right]^{-2}$ ($\kappa = 0.384$, $C = 3.682$), one obtains a slightly lower value ($C_f = 0.0043$). The 10\% difference can be ascribed to the Coles--Fernholz relation holding for a fully developed TBL along a smooth wall.

\begin{figure}[htbp]
\centering
\includegraphics[width=\textwidth]{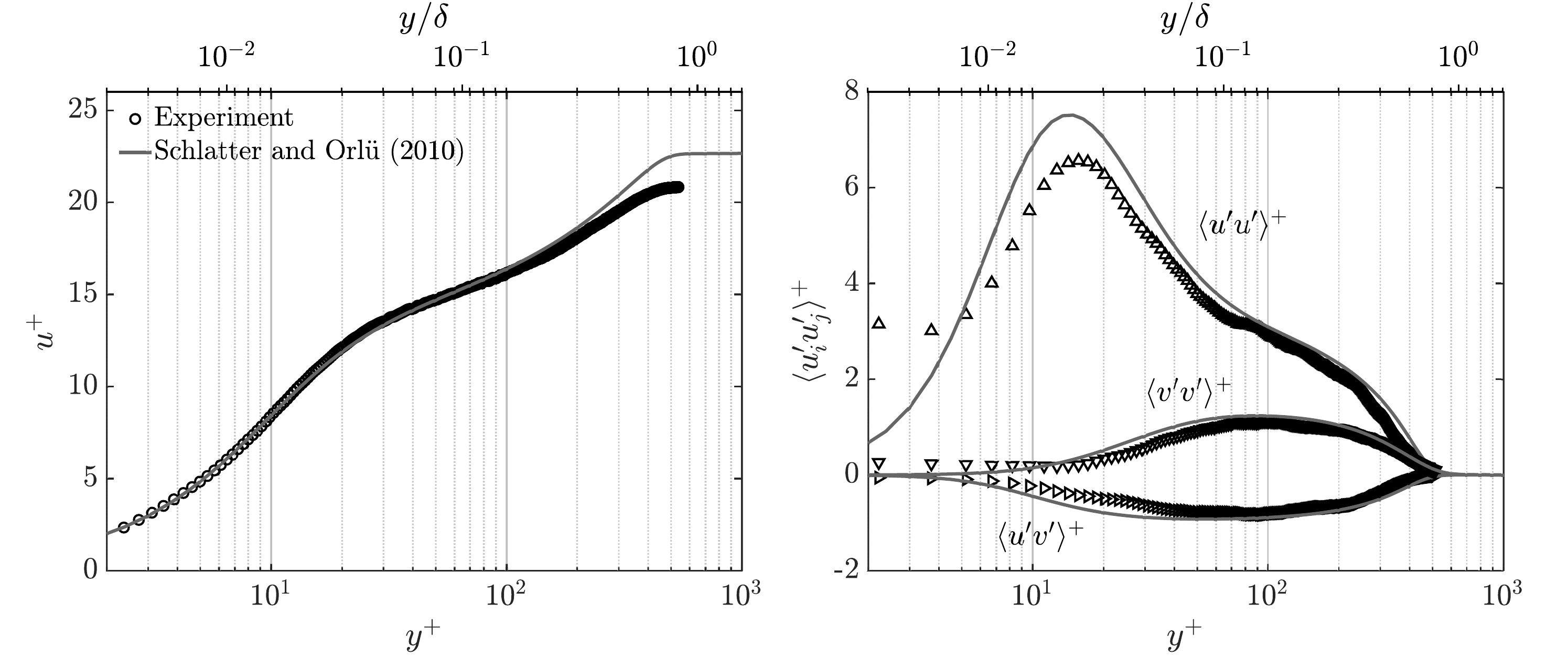}
\caption{Wall-normal profiles of velocity statistics scaled in wall units at $x = \qty{0}{\milli\metre}$. Left: mean velocity profile. Right: ensemble-averaged Reynolds stresses.}\label{fig:11}
\end{figure}

Figure~\ref{fig:11} depicts profiles of mean streamwise velocity (left) and Reynolds stresses (right). Ensemble and streamwise averaging over $1.5\delta$ is applied to improve statistical convergence and hereinafter indicated by the operator $\langle \cdots \rangle$. Comparison with DNS data at a comparable $Re_\theta$ \citep{schlatter2010} indicates a good agreement, except for the reduced wake region attributed to residual effects of the tripping device \citep{sanmiguelvila2017}. However, the viscous sublayer $2 \leq y^+ \leq 5$ is well captured, vouching for an accurate wall shear stress measurement and a negligible influence of non-canonical features in the outer region of the boundary layer.

Due to the larger interrogation windows partly overlapping the wall, the streamwise normal Reynolds stress features a discrepancy in the characteristic peak value and an overprediction in the near-wall region ($y^+ < 4$). This modulation is also visible in the behaviour of $\langle u'v'\rangle$ and $\langle v'v'\rangle$.

\subsection{Actuated regime}\label{subsec:actuated}

Spanwise wall forcing is imposed at $T^+ \approx 101$ and $A^+ \approx 7$, a combination of parameters within the mechanical limitations of the apparatus, yet delivering an observable drag reduction according to \cite{kempaiah2020}. The maximum achievable value of $A^+$ is set (as discussed in Sect.~\ref{sec:intro}, experiments allow fixing the peak-to-peak displacement $2D_m$), whereas a near-optimum value of $T^+$ was chosen.

\begin{figure}[htbp]
\centering
\includegraphics[width=\textwidth]{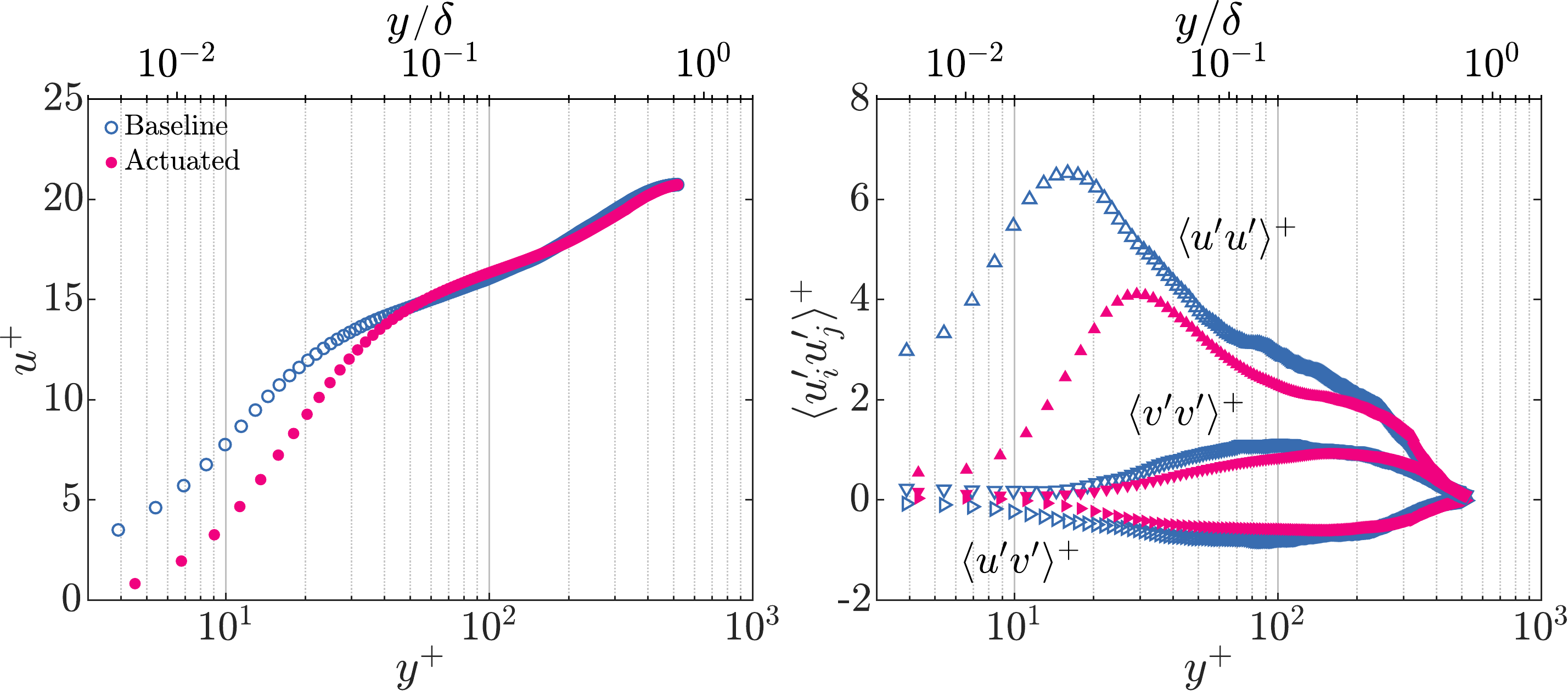}
\caption{Left: inner-scaled mean streamwise velocity profile on the stationary and oscillating wall. Right: inner-scaled ensemble-averaged Reynolds stresses.}\label{fig:12}
\end{figure}

Figure~\ref{fig:12} depicts the comparison between non-actuated and actuated statistics. Profiles have been averaged at $5.5 \leq x/\delta \leq 6.5$, where all transient effects except the shear stress peak have ceased and the full-fledged drag-reduced state is reached (see Sect.~\ref{subsec:spatial_stats} on spatial development).

A drag reduction based on the streamwise- and time-averaged shear stress of $\mathcal{R} = 32\%$ is found, in good agreement with the value reported by \cite{ricco2008} at slightly different parameters and fixed $D_m^+$.

The reduced slope of $u^+$ in the viscous sublayer is visible, and clearly indicative of a drag-reduced state. When the reference friction velocity is used to scale the quantities, the usual attenuation of $\langle u'u'\rangle^+$ is visible up to $y^+ \approx 150$, indicating an alteration of the character of the turbulent flow up to the buffer layer. The peak of $\langle u'u'\rangle$ is reduced by 37\% and shifted by about $13\delta_\nu$ away from the wall. These results indicate that streamwise velocity fluctuations near the wall are reduced to a greater extent, proportionally, than the corresponding mean flow \citep{riccowu2004}.

A similar behaviour in terms of near-wall suppression and peak shift, although less evident, is spotted for the wall-normal and shear stresses. As for $\langle u'u'\rangle$, actuation reduces the peak of $\langle u'v'\rangle^+$ by about 30\% and shifts it by $20\delta_\nu$ away from the wall. When scaled in the drag-reduced units, reference and actuated profiles (not shown here) show a good match beyond $y^* \approx 20$ \citep{ricco2021}.

\subsection{Skin friction}\label{subsec:skin_friction}

\begin{figure}[htbp]
\centering
\includegraphics[width=\textwidth]{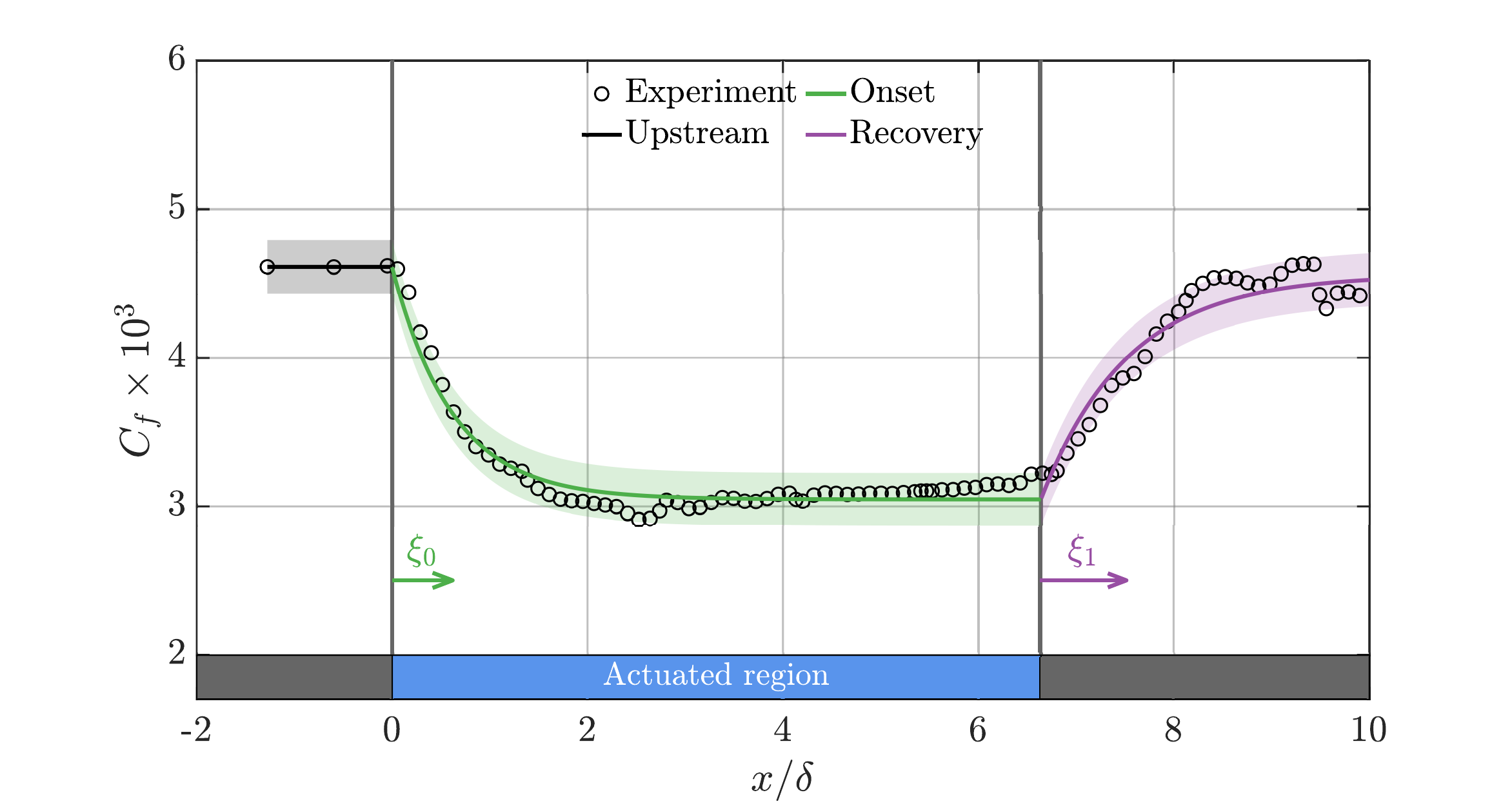}
\caption{Trend of $C_f(x)$ for the onset (green) and recovery (purple) transient. Points represent experimental measurements, which have been downsampled by approximately $1/50$ to enhance legibility; solid line is the exponential fit. Vertical solid lines mark the beginning and end of the actuation. Green (purple) arrow indicates the onset (recovery) adaptation length. Shaded regions represent prediction intervals at 95\% confidence level.}\label{fig:13}
\end{figure}

Figure~\ref{fig:13} depicts the spatial development of friction for the controlled case, alongside the exponential fit, represented in green for the onset and purple for the recovery transient, respectively. The model is defined as a piecewise function $f$ of the following form:
\begin{equation}\label{eq:10}
f(x) = \begin{cases}
f_0, & x \leq x_0 \\[1ex]
f_0 - \Delta f\left(1 - \mathrm{e}^{-\frac{x - x_0}{\xi_0}}\right), & x_0 < x \leq x_1 \\[1ex]
C + (f_r - C)\left(1 - \mathrm{e}^{-\frac{x - x_1}{\xi_1}}\right), & x > x_1
\end{cases}
\end{equation}
Consistently with the nomenclature introduced in Eqs.~(\ref{eq:3}) and (\ref{eq:6}), $f_0$ and $f_r$ indicate the baseline values pre- and post-actuation, and $\Delta f = f_0 - f_c$. $C = f_0 - \Delta f\left(1 - \mathrm{e}^{-(x_1 - x_0)/\xi_0}\right)$ is the exact value at the end of the oscillating wall, ensuring the function remains continuous.

A total of five parameters $(f_0, f_r, \Delta f, \xi_0, \xi_1)$ are fitted to the experimental data by means of a least-squares optimisation. Before the fitting procedure, outliers are removed by applying a sliding median filter.

Uncertainty on the fitted parameters is quantified by determining the 95\% confidence interval from the standard error of the fit. It is scaled using the Student's $t$-distribution evaluated for the degrees of freedom of the fit. These correspond to the number of uncorrelated datapoints along $x$, one every four due to the 75\% overlap of the interrogation windows, minus the order of the fit, i.e.\ five.

The exponential fit yields $\xi_0 \sim 0.59\delta \pm 0.02\delta$ for the onset transient and $\xi_1 \sim 0.88\delta \pm 0.03\delta$ for the recovery one. The tight uncertainty is owed to the large number of degrees of freedom and decreases as $1/\sqrt{N_{\mathrm{dof}}}$; to retain the natural scatter of the data one must consider the prediction interval \citep{seber1989}, i.e.\ the probability that a future observation falls within a certain range. The authors therefore choose to report the covariance-based uncertainty at $2\sigma$ and plot the prediction intervals as shaded bands.

Fitting the trends of Fig.~\ref{fig:1} with the same model, one can compare the present adaptation lengths with the literature. The results are collected in Table~\ref{tab:5}.

\begin{table}[htbp]
\caption{Comparison of adaptation lengths estimated with exponential fit for present study and literature data. RW2004: \cite{riccowu2004}; LL2013: \cite{lardeau2013}}\label{tab:5}%
\begin{tabular}{@{}lcccc@{}}
\toprule
Reference & $\xi_0/\delta$ & $\sigma_{\xi_0}/\delta$ & $\xi_1/\delta$ & $\sigma_{\xi_1}/\delta$ \\
\midrule
Present & 0.59 & 0.02 & 0.88 & 0.03 \\
RW2004  & 0.73 & 0.14 & 0.83 & 0.24 \\
LL2013  & 0.45 & 0.05 & 0.53 & 0.05 \\
\botrule
\end{tabular}
\end{table}

With this model, the recovery transient is consistently longer than the onset one, beyond the standard error. The slight differences in the value of $\xi$ may stem from different implementations of the actuation mechanism between experiments and numerical works. For instance, \cite{lardeau2013} utilise a hyperbolic-tangent ramp-up over $0.2\delta$, which may affect the estimation of the adaptation length. Conversely, experiments suffer from limited tolerances at plate junctions, bearings and imposition of the control law.

The recovery transient is especially relevant to the application of this technique by discrete actuation or even passive surrogates thereof: a longer effect of drag reduction past the actuator allows a more prolonged exploitation of the technique before a subsequent actuator is needed.

\subsection{Spatial development of turbulence statistics}\label{subsec:spatial_stats}

The evolution of Reynolds stresses across the entire domain of interest is presented here, both in the form of two-dimensional contours and streamwise profiles. All statistics, as well as the wall-normal distance, are normalised by the reference friction velocity to appreciate the differences between the baseline and actuated cases \citep{agostini2014}. To enhance legibility, the data are spatially low-pass filtered along the streamwise direction by a Gaussian filter with a \qty{1}{\milli\metre}-wide ($0.02\delta$) kernel.

\subsubsection{Streamwise Reynolds stress}\label{subsubsec:uu}

\begin{figure}[htbp]
\centering
\includegraphics[width=\textwidth]{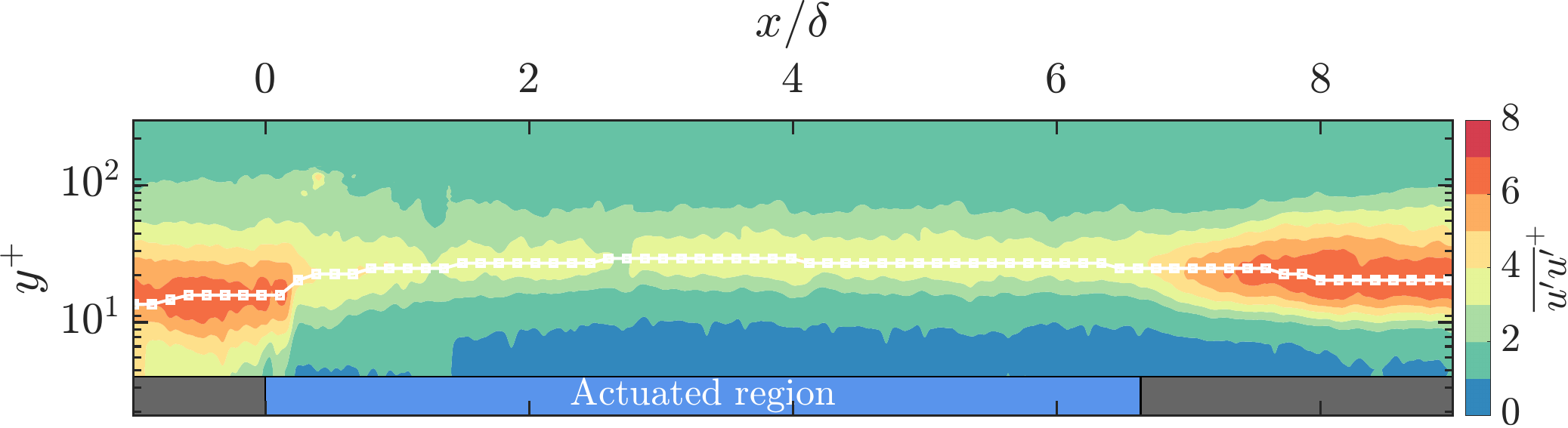}
\caption{Spatial development of streamwise Reynolds stress $\overline{u'u'}^+$ along the streamwise direction for the actuated case. The patched rectangle represents the behaviour of the wall, blue where control is active, grey where control is inactive. White markers represent the position of the peak (downsampled to 1:40).}\label{fig:14}
\end{figure}

Contours of the streamwise normal Reynolds stress scalar field in the actuated case are depicted in Fig.~\ref{fig:14}, employing a logarithmic scale along the wall-normal direction.

The distinct energetic peak at $y^+ \approx 15$, visible in Fig.~\ref{fig:11}, is heavily dampened already within the first $\delta$. This reflects a general decline in turbulence activity, here representative of streak intensity. The latter is inherently linked to the shear stress production and ultimately to skin friction \citep{jimenez1999, kim2011}. A reduction of approximately 40\% compared to the baseline figure is observed, corroborating the trends shown in the profiles of Fig.~\ref{fig:12}. The wall-normal shift of the peak is also evident in its streamwise-developing fashion, settling to the drag-reduced value in $1 \div 2\delta$.

The recovery, taking place from the trailing edge of the actuator ($x/\delta \approx 6.5$), occurs over a streamwise extent of $\sim 1\delta$, after which the streamwise stress recovers to the baseline value.

\begin{figure}[htbp]
\centering
\includegraphics[width=\textwidth]{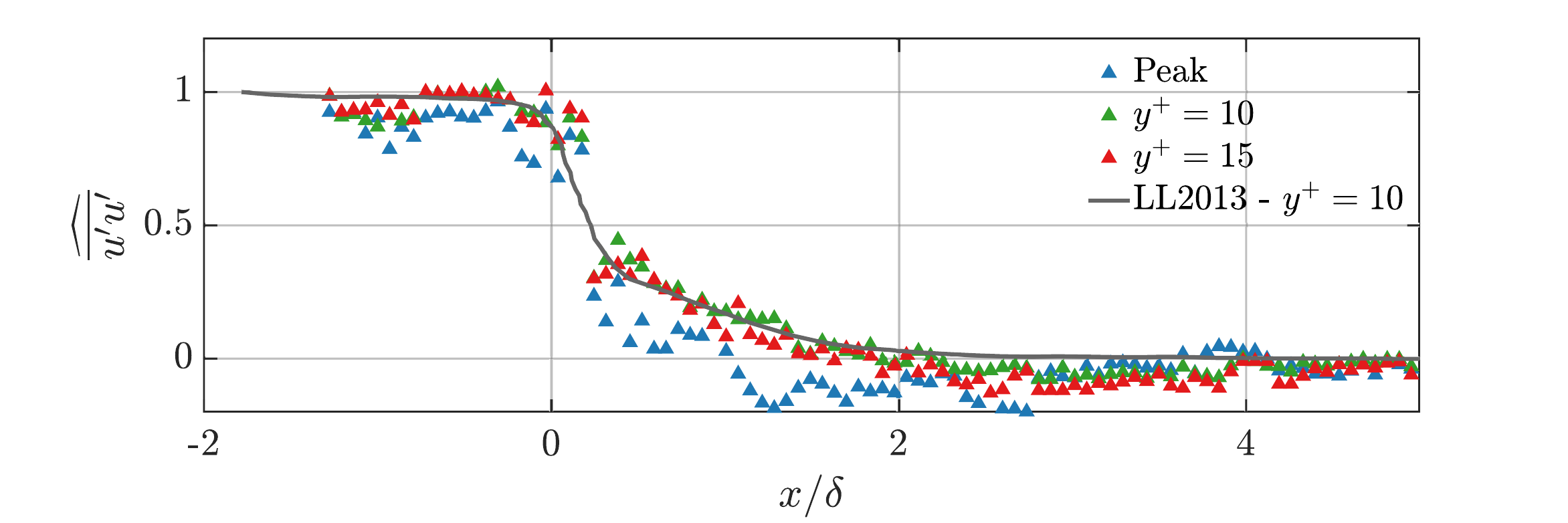}
\caption{Normalised streamwise stress compared with trend from \cite{lardeau2013}.}\label{fig:15}
\end{figure}

To better gauge the extent of the transient, the peak value and streamwise profiles at $y^+ = 10$ and $y^+ = 15$ are extracted from the contours. The choice of the wall-normal heights is dictated by the observation that the near-wall streaks are most pronounced in this range \citep{touber2012}, where the peak of turbulence production also resides. Figure~\ref{fig:15} displays the spatial evolution of said profiles, compared to the trend from \cite{lardeau2013} at $y^+ = 10$. To account for the discrepancy in $\mathcal{R}$ and given that the focus of the study is the streamwise adaptation rather than drag reduction performance, the curves have been normalised to allow direct comparison. To this end, the following transformation is applied:
\begin{equation}\label{eq:11}
\hat{f} = \frac{f - f_c}{\Delta f} = 1 + \frac{f - f_0}{\Delta f}
\end{equation}

\begin{figure}[htbp]
\centering
\includegraphics[width=\textwidth]{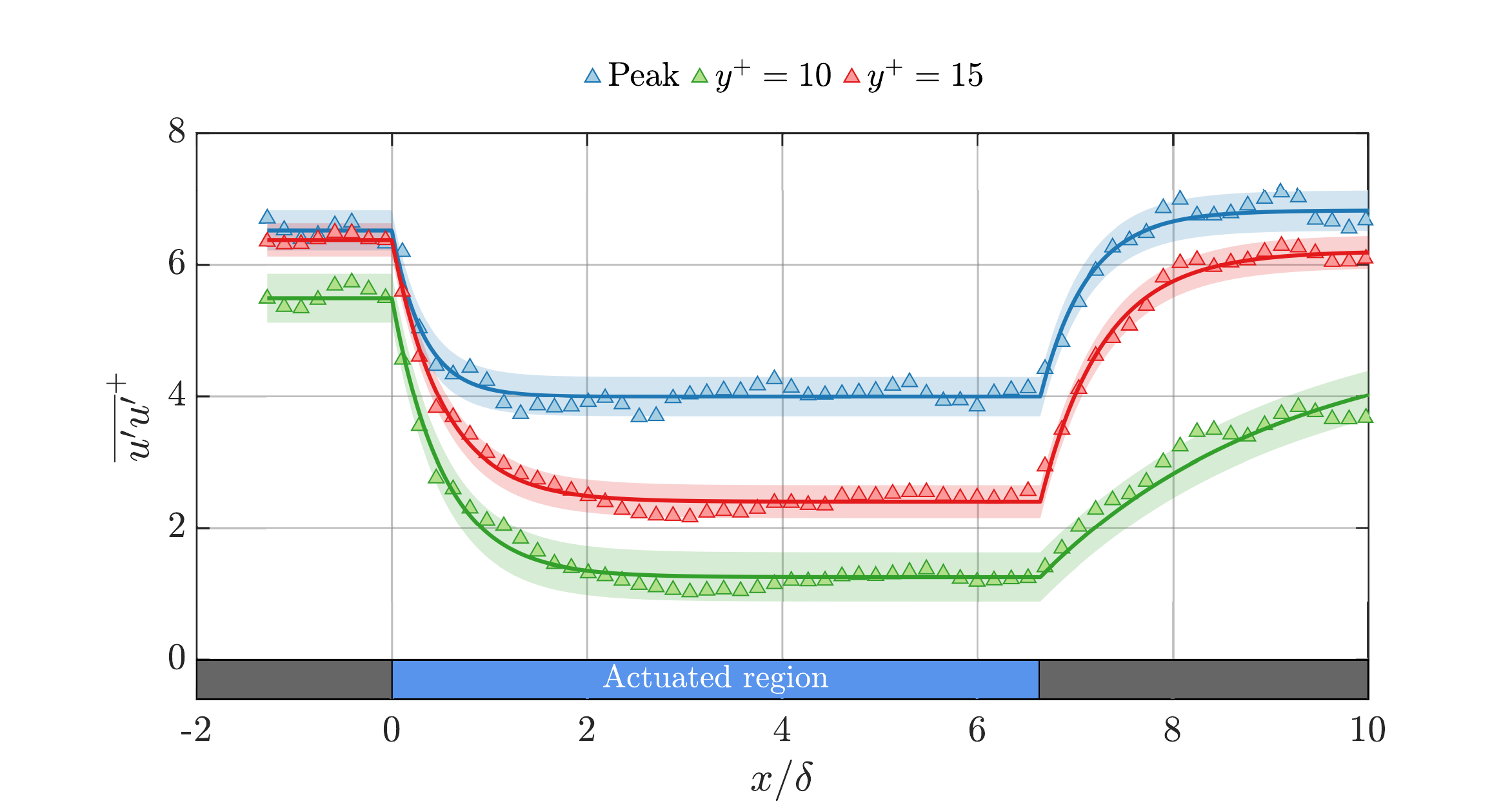}
\caption{Streamwise evolution of profiles of $\overline{u'u'}^+$ (symbols, 1:50 downsampling), alongside the corresponding exponential fits (solid lines). Shaded regions represent prediction intervals at 95\% confidence level.}\label{fig:16}
\end{figure}

When normalised, the profiles come to an overlap with each other and agree with previous literature (solid line). An analogous fitting procedure is applied to the spatial development of the streamwise velocity fluctuations, and the result is depicted in Fig.~\ref{fig:16}.

Firstly, it is worth noting that, since the peak of $\overline{u'u'}$ in the unperturbed flow resides at $y^+ \approx 15$, the two profiles start from the same value at the oscillation onset. However, as the flow advances into the actuated region, the two start to deviate from one another. This is because, when extracting the profile at a constant (nominal) $y^+$ height, two physical processes are highlighted at the same time. One is indeed the dampening of the turbulence peak; the other is the thickening of the viscous sublayer, which entails the wall-normal shift of the peak. Because in general these two effects have different scales, as also evident from Fig.~\ref{fig:14}, the estimation of the transient is also biased.

The adaptation lengths for the streamwise profiles of $\overline{u'u'}$ are collected in Table~\ref{tab:6}: comparing them to Table~\ref{tab:5}, the streamwise stress exhibits a faster response in both transients compared to skin friction.

\begin{table}[htbp]
\caption{Comparison of adaptation lengths of streamwise stress profiles for the present study and literature.}\label{tab:6}%
\begin{tabular}{@{}llcccc@{}}
\toprule
\multicolumn{2}{@{}l}{Quantity} & $\xi_0/\delta$ & $\sigma_{\xi_0}/\delta$ & $\xi_1/\delta$ & $\sigma_{\xi_1}/\delta$ \\
\midrule
Peak & & 0.33 & 0.02 & 0.48 & 0.02 \\
$y^+ = 15$ & & 0.53 & 0.01 & 0.64 & 0.02 \\
\multirow{2}{*}{$y^+ = 10$} & Present & 0.53 & 0.02 & 2.50 & 0.21 \\
 & LL2013 & 0.37 & 0.03 & -- & -- \\
\botrule
\end{tabular}
\end{table}

Yet, owing to the combination of effects described above, the profile at $y^+ = 15$ displays a greater $\xi$ compared to the peak. This effect is even more noticeable in the recovery at $y^+ = 10$. Inspecting Fig.~\ref{fig:14} once more, the re-establishment of the canonical structure of the boundary layer past the actuated segment is more gradual compared to the sharp onset. This is a plausible explanation for the recovery extending for a wider streamwise fetch, which is also confirmed here; however, when departing too far from the peak, this effect is exacerbated and reflects in $\xi_1 \gg \xi_0$. Therefore, in favour of a reliable and systematic assessment of the turbulence response, the peak is chosen as the meaningful quantity to track.

\subsubsection{Reynolds shear stress}\label{subsubsec:uv}

\begin{figure}[htbp]
\centering
\includegraphics[width=\textwidth]{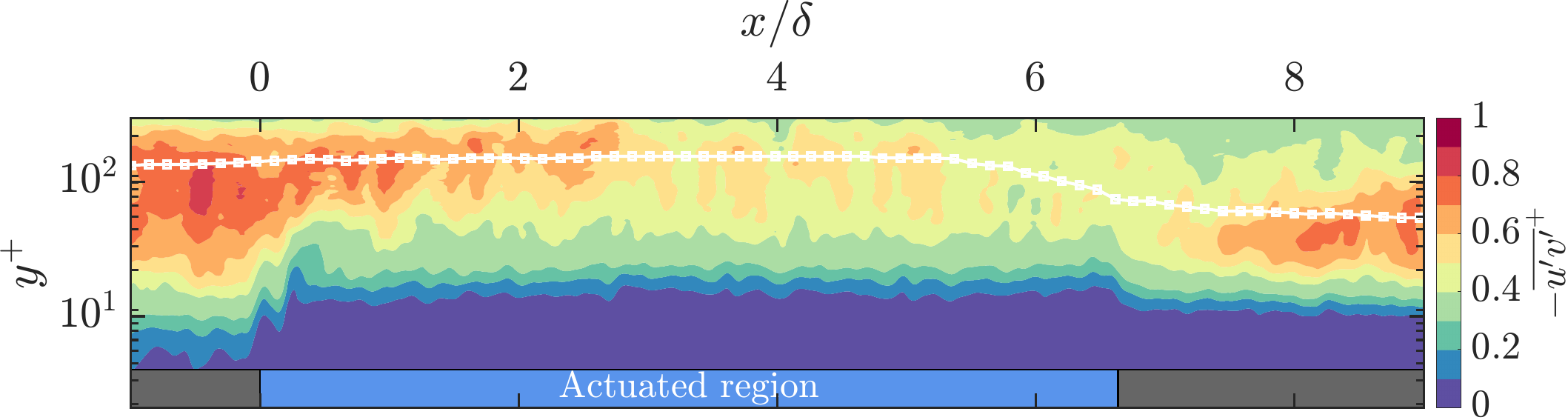}
\caption{Spatial development of Reynolds shear stress $-\overline{u'v'}^+$ along the streamwise direction for the actuated case. The same conventions as in Fig.~\ref{fig:14} are followed.}\label{fig:17}
\end{figure}

The Reynolds shear stress evolution under spanwise wall oscillations is displayed in Fig.~\ref{fig:17}. Both the peak attenuation and wall-normal shift are visible from the contours, with a decrease of roughly 30\% that however extends for $\approx 3\delta$, hinting at a longer adaptation process.

\begin{figure}[htbp]
\centering
\includegraphics[width=\textwidth]{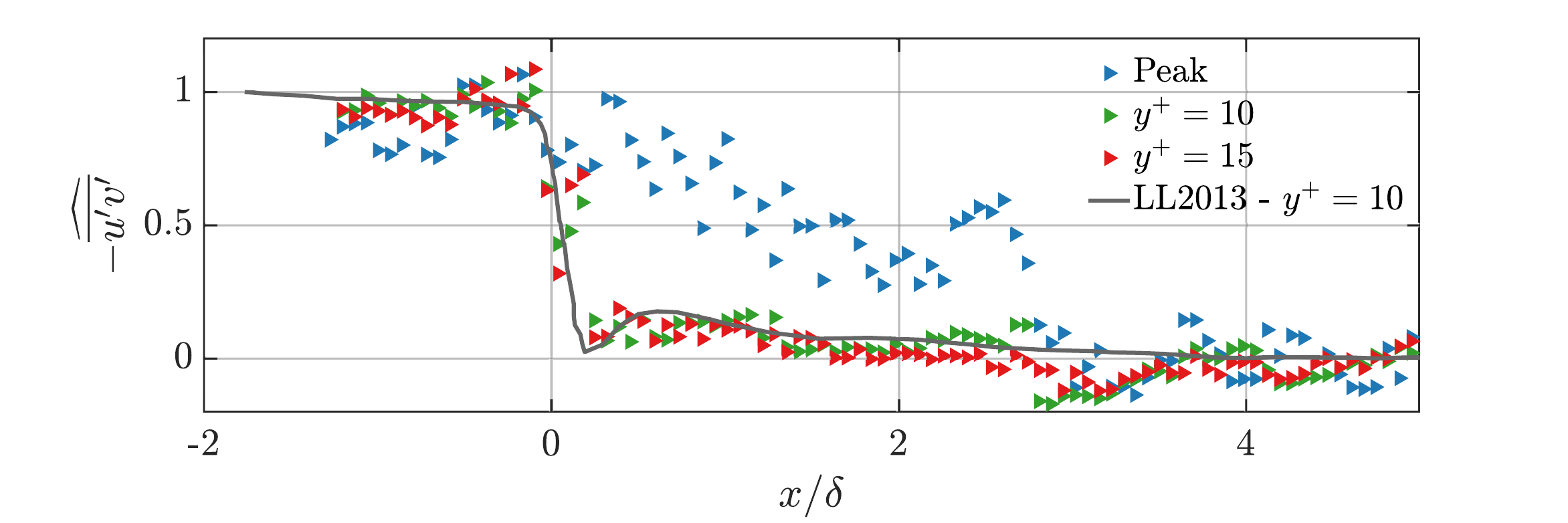}
\caption{Profiles of normalised Reynolds shear stress compared with trend from \cite{lardeau2013}.}\label{fig:18}
\end{figure}

While at $y^+ = 10$ and $15$ the profiles closely follow the trend of \cite{lardeau2013}, with an abrupt drop within $0.2\delta$, the peak presents a more gradual decrease. Since the latter is shifted to $y^+ \approx 100$, the extraction of a profile at $y^+ = 10 \div 15$ only captures the effect of the thicker viscous sublayer, which, as mentioned above, manifests a sudden response to the wall actuation. Normalised profiles of the peak and values at $y^+ = 10$ and $15$ for the onset transient are represented in Fig.~\ref{fig:18} against literature data.

\begin{figure}[htbp]
\centering
\includegraphics[width=\textwidth]{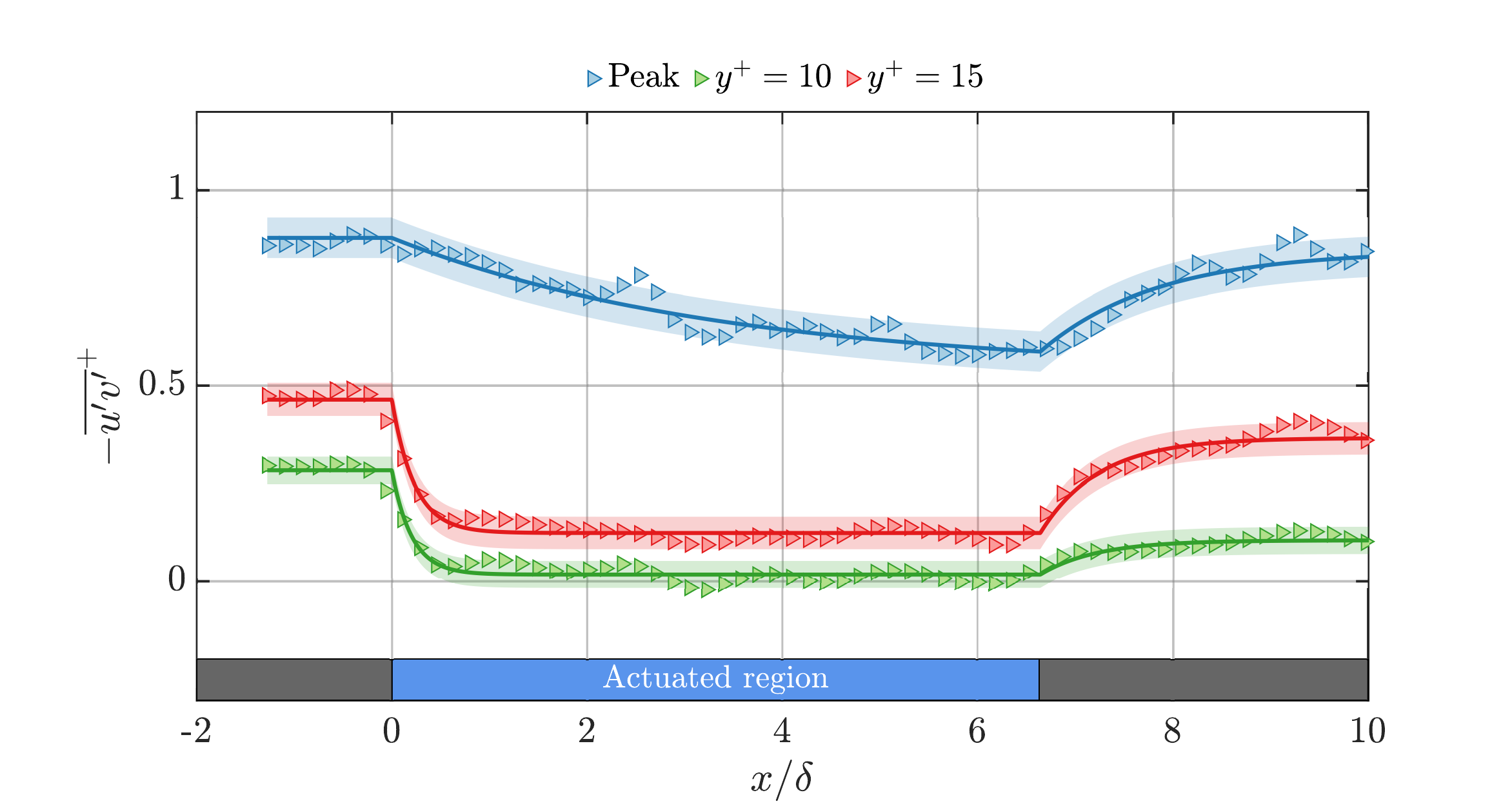}
\caption{Streamwise evolution of profiles of $-\overline{u'v'}^+$ (symbols, 1:50 downsampling), alongside the corresponding exponential fits (solid lines). Shaded regions represent prediction intervals at 95\% confidence level.}\label{fig:19}
\end{figure}

The exponential fit described by Eq.~(\ref{eq:10}) yields the curves depicted in Fig.~\ref{fig:19}.

The slower onset transient for the peak can be appreciated and is quantitatively mirrored in the higher value of $\xi$ reported in Table~\ref{tab:7}.

\begin{table}[htbp]
\caption{Comparison of adaptation lengths of Reynolds shear stress profiles for the present study and literature}\label{tab:7}%
\begin{tabular}{@{}llcccc@{}}
\toprule
\multicolumn{2}{@{}l}{Quantity} & $\xi_0/\delta$ & $\sigma_{\xi_0}/\delta$ & $\xi_1/\delta$ & $\sigma_{\xi_1}/\delta$ \\
\midrule
Peak & & 5.65 & 0.77 & 1.09 & 0.09 \\
$y^+ = 15$ & & 0.23 & 0.01 & 0.60 & 0.04 \\
\multirow{2}{*}{$y^+ = 10$} & Present & 0.20 & 0.01 & 0.58 & 0.09 \\
 & LL2013 & 0.07 & 0.01 & -- & -- \\
\botrule
\end{tabular}
\end{table}

Adaptation lengths of the shear stress at $y^+ = 10$ and $15$ are significantly shorter and about double the value from \cite{lardeau2013}. Once again, the $0.2\delta$ ramp in their study and the mechanical tolerances in the experimental realisation may play a role. The solid line in Fig.~\ref{fig:18} presents a sudden dip at $x/\delta = 0 \div 0.2$, followed by a more gradual decrease; manually fitting the exponential to filter out the dip gives $\xi_0 \sim 0.40\delta$. Interestingly, both profiles at $y^+ = 10$ and $y^+ = 15$ present $\xi_1/\xi_0 \approx 3$. This ratio is significantly larger than the one found for $C_f$ and $\overline{u'u'}$, which typically settles to $1.2 \div 1.4$ for the present experiment as well as available literature. This observation further elucidates the difference in the response of the two aforementioned effects of drag reduction; still, this claim is subject to uncertainties due to limits of PIV in estimating second-order moments, especially for large FOVs and limited temporal fidelity.

\subsection{Quadrant analysis}\label{subsec:quadrant}

Quadrant analysis is employed to link the behaviour of statistics under the control action to the phenomenology of the boundary layer. Specifically, this type of analysis is aimed at understanding the physics of the generation of Reynolds shear stress and TKE production \citep{wallace2016}.

It is established that spanwise wall oscillations impact quasi-streamwise vortices, primarily suppressing Q2 events (i.e.\ ejections) at near-optimal oscillation periods and, to a lesser extent, Q4 (i.e.\ sweeps) \citep[see for instance][]{yakeno2014}. In fully developed conditions, mass conservation implies that ejections and sweeps match each other, at least in an average fashion.

However, it is not yet clear whether there is a difference in adaptation length of the response between the two. \cite{lardeau2013} evaluated the joint PDF at $y^+ = 10$, at the oscillation onset and at $x/\delta = 1$. They observe a shift of the maximum to the right-hand side between the reference and actuated cases, ascribed to the preferred decline in ejections. Yet, between the two actuated cases at different $x$ the only difference spotted is a reduced correlation between $u'$ and $v'$, thus a reduction in shear stress. It therefore remains unclear whether the oscillating wall, in a transient sense, imparts the control preferentially on a specific mechanism.

\begin{figure}[htbp]
\centering
\includegraphics[width=\textwidth]{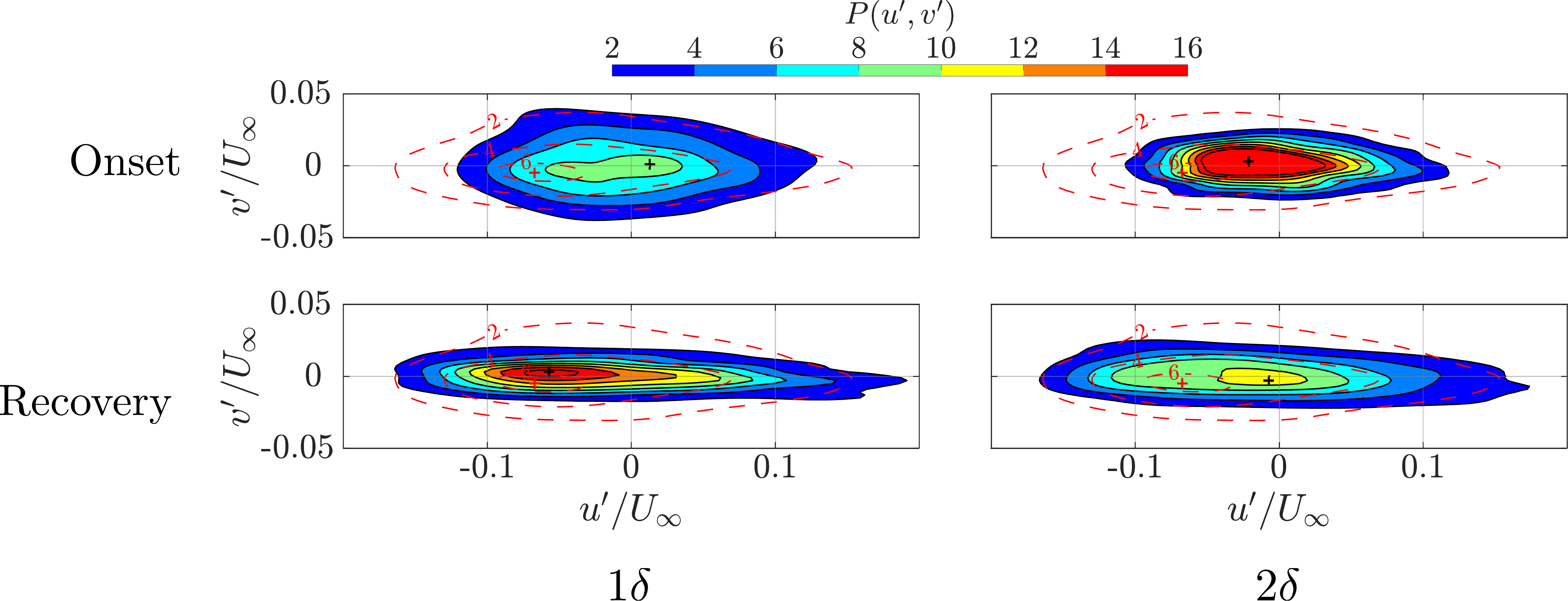}
\caption{Contours of the joint PDF $P(u', v')$ of $u'$ and $v'$ for the onset (top row) and recovery (bottom row) at $x/\delta = 1$ (left column) and $2$ (right column). Dashed lines represent contours of $P(u', v')$ in the baseline case.}\label{fig:20}
\end{figure}

The joint PDF $P(u', v')$ is evaluated at $y^+ = 10$, at $1\delta$ and $2\delta$ past the onset and termination of the oscillations, respectively. Contours of $P(u', v')$ are depicted in Fig.~\ref{fig:20}, overlaid with the ones of the baseline case.

The evolution of the joint PDF along the onset transient is characterised by a shift of the peak from the left-hand side ($u' < 0$) to the centre. This highlights the main action of control on the low-speed streaks, alongside the shrinking of the PDF and sharpening of the peak.

Such reduction in breadth happens first along the $x$-axis and then on the $y$-axis, indicating a slower response of $v'$-fluctuations. This observation is consistent with the longer transient of $\overline{u'v'}$ as opposed to $\overline{u'u'}$. The longer recovery transient is corroborated by the observation that at $2\delta$ past the rearmost edge of the oscillating plate the joint PDF is still recovering from the drag-reduced state. Once again, the $v'$-fluctuations persist in the actuated state for a longer streamwise extent.

\begin{figure}[htbp]
\centering
\includegraphics[width=\textwidth]{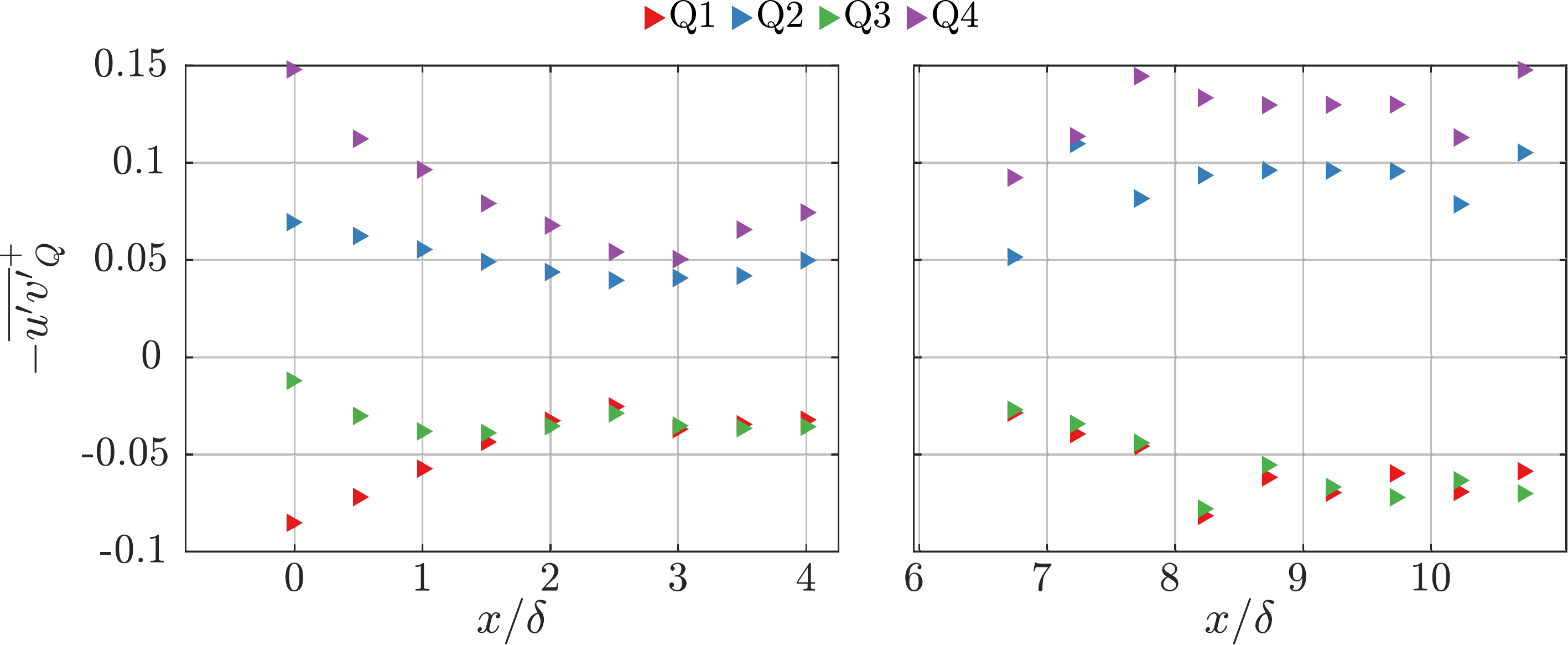}
\caption{Quadrant contributions to $-\overline{u'v'}_Q^+$ at $y^+ = 10$ for onset (left) and recovery (right).}\label{fig:21}
\end{figure}

From this type of analysis, however, it is not yet clear whether a certain type of motion is preferentially impacted by the wall oscillations. To further discriminate the impact of spanwise wall forcing on turbulence-producing events, the Reynolds shear stress is computed as the covariance of the velocity fluctuations:
\begin{equation}\label{eq:12}
\overline{u'v'} = \int_{-\infty}^{+\infty}\int_{-\infty}^{+\infty} u'v' P(u', v')\,\mathrm{d}u'\,\mathrm{d}v'
\end{equation}
In this way, the contribution of the separate quadrants to the shear stress can be isolated by appropriately conditioning the integrand. Figure~\ref{fig:21} displays the resultant trend for the two transients, assessed every $0.5\delta$ until $4\delta$ downstream of the start (cessation) of the actuated segment.

The response of shear stress elucidated in Fig.~\ref{fig:19} is here decomposed by quadrant and shows that ejections are primarily responding to the control action, being damped by a factor of 3 after $1\delta$. Comparatively, Q4 contributions degrade slightly more gradually.

Interestingly, a slightly increasing trend is observed after $3\delta$, consistently with the long-term trend of the skin friction (Fig.~\ref{fig:13}). The recovery after the actuator (Fig.~\ref{fig:21}, right) features rising values of Q2 and Q4 components, with the latter attesting to higher levels after $2\delta$, suggesting that turbulent sweeps are activated more rapidly than the ejections.

These observations are consistent with the credited theory of primary suppression of ejections, established in full-fledged drag-reduced \citep{yakeno2014} and now also in streamwise-developing conditions. This is ultimately related to the inhibition of the hairpin autogeneration cycle \citep{kempaiah2020}, which is primarily linked to ejections and then, by continuity, entails a drop in sweeping motions, too. On the other hand, the skin friction recovery is ascribed to the enhancement of Q4 occurrences.

\section{Conclusions}\label{sec:conclusions}

The present study investigates the transient response of a turbulent boundary layer subjected to spanwise wall oscillations using PIV. The primary focus is on quantifying the spatial development of turbulence statistics and wall shear stress following the onset of actuation and past its ending, and on modelling the friction response with a single exponential function.

A problem-specific approach is proposed to estimate the wall shear stress along an extended domain streamwise, compromising the spatial resolution of the measurements. An arctangent fit of the near-wall velocity profile is adopted, which enables the use of the measured data beyond the linear region. The error analysis and comparison between the linear and arctangent fits reveal the opportunity to further decrease the bias error by a combination of the two. An assessment of the method returns a wide plateau of error below 5\% over a range of choices for the gap height and the fitting range ($[y_l^+, y_u^+] < [10, 30]$).

Wall oscillations are imposed at $T^+ \approx 101$ and $A^+ \approx 7$, yielding a drag reduction $\mathcal{R} = 32\%$, in close agreement with literature data.

The wall shear stress evolution is characterised via the friction coefficient $C_f(x)$ computed from locally averaged velocity profiles. Following the onset of oscillations, $C_f$ shows a rapid drop, and its streamwise variation is modelled with a negative exponential function. The functional form fits well the onset and the recovery transients. The response length is determined for onset and recovery, yielding $\xi_0 \sim 0.6\delta$ and $\xi_1 \sim 0.9\delta$, respectively.

The spatial evolution of second-order statistics is assessed by inspecting their inner-scaled contours and profiles over the whole region of interest. Their behaviour also follows an exponential-type evolution for onset and recovery. The transient length for recovery appears to be longer than that at the onset. However, this condition is reversed for the peak of the turbulent shear stress. This may be ascribed to the fact that the recovery converges to a value lower than that upstream of the oscillations, as well as a significant covariance between the asymptotic value and relaxation length.

The quadrant analysis identifies a more rapid response of Q2 events at the onset. The situation reverses at recovery, where Q4 events drive the increase of skin friction past the oscillating plate.

\backmatter

\bmhead{Acknowledgements}
The authors wish to thank the technical staff of the Aerodynamics laboratories at TU Delft. Specifically, the help of F. Donker-Duyvis, D. Bruikman and G.-J. Berends is gratefully acknowledged.

\section*{Declarations}

\begin{description}
\item[Funding] This work was financially supported by the Netherlands Enterprise Agency under grant number TSH21002.
\item[Conflict of interest] The authors declare no conflict of interest.
\item \textbf{Data availability:} A full dataset accompanying the individual figures is available online and can be found at \url{https://www.doi.org/10.4121/659f9160-21f0-4d7a-bdba-705658558fda}.
\item[Author contributions] MG, FSchr and FS contributed to the study conception and design. Investigation and formal analysis were performed by MG and NS; MG carried out methodology, visualisation, data curation and wrote the original draft. All authors contributed to review and editing of the manuscript. FSchr and FS worked in supervision and helped in funding acquisition.
\end{description}

\bmhead{Open Access}
This article is licensed under a Creative Commons Attribution 4.0 International License, which permits use, sharing, adaptation, distribution and reproduction in any medium or format, as long as you give appropriate credit to the original author(s) and the source, provide a link to the Creative Commons licence, and indicate if changes were made. The images or other third-party material in this article are included in the article's Creative Commons licence, unless indicated otherwise in a credit line to the material. If material is not included in the article's Creative Commons licence and your intended use is not permitted by statutory regulation or exceeds the permitted use, you will need to obtain permission directly from the copyright holder. To view a copy of this licence, visit \url{http://creativecommons.org/licenses/by/4.0/}.

\bibliography{sn-bibliography}

\begin{thebibliography}{40}
\providecommand{\natexlab}[1]{#1}
\providecommand{\url}[1]{{#1}}
\providecommand{\urlprefix}{URL }
\providecommand{\doi}[1]{\url{https://doi.org/#1}}
\providecommand{\eprint}[2][]{\url{#2}}
 \bibcommenthead

\bibitem[{Agostini et~al.(2014)Agostini, Touber, and Leschziner}]{agostini2014}
Agostini L, Touber E, Leschziner MA (2014) Spanwise oscillatory wall motion in channel flow: drag-reduction mechanisms inferred from {DNS}-predicted phase-wise property variations at {$Re_\tau = 1000$}. J Fluid Mech 743:606--635. \doi{10.1017/jfm.2014.40}

\bibitem[{Bogard et~al.(2000)Bogard, Ball, and Wassen}]{bogard2000}
Bogard D, Ball K, Wassen E (2000) Drag reduction for turbulent boundary layer flows using an oscillating wall. Tech. rep., Mechanical Engineering Department, The University of Texas at Austin, Austin, TX

\bibitem[{Chauhan et~al.(2009)Chauhan, Monkewitz, and Nagib}]{chauhan2009}
Chauhan KA, Monkewitz PA, Nagib HM (2009) Criteria for assessing experiments in zero pressure gradient boundary layers. Fluid Dyn Res 41(2):021404. \doi{10.1088/0169-5983/41/2/021404}

\bibitem[{Choi et~al.(1998)Choi, DeBisschop, and Clayton}]{choi1998}
Choi KS, DeBisschop JR, Clayton BR (1998) Turbulent boundary-layer control by means of spanwise-wall oscillation. AIAA J 36(7):1157--1163. \doi{10.2514/2.526}

\bibitem[{Dacome et~al.(2024)Dacome, M{\"o}rsch, Kotsonis, and Baars}]{dacome2024}
Dacome G, M{\"o}rsch R, Kotsonis M, et~al (2024) Opposition flow control for reducing skin-friction drag of a turbulent boundary layer. Phys Rev Fluids 9:064602. \doi{10.1103/physrevfluids.9.064602}

\bibitem[{Di~Cicca et~al.(2002)Di~Cicca, Iuso, Spazzini, and Onorato}]{dicicca2002}
Di~Cicca GM, Iuso G, Spazzini PG, et~al (2002) Particle image velocimetry investigation of a turbulent boundary layer manipulated by spanwise wall oscillations. J Fluid Mech 467:41--56. \doi{10.1017/S002211200200157X}

\bibitem[{Faleiros et~al.(2019)Faleiros, Tuinstra, Sciacchitano, and Scarano}]{faleiros2019}
Faleiros DE, Tuinstra M, Sciacchitano A, et~al (2019) Generation and control of helium-filled soap bubbles for {PIV}. Exp Fluids 60:40. \doi{10.1007/s00348-019-2687-4}

\bibitem[{Hendriksen et~al.(2025)Hendriksen, Sciacchitano, and Scarano}]{hendriksen2025}
Hendriksen LA, Sciacchitano A, Scarano F (2025) Skin friction topology on generic objects from {3D} particle tracking. In: Proceedings of the 16th International Symposium on Particle Image Velocimetry (ISPIV), Tokyo, Japan

\bibitem[{Jim{\'e}nez and Pinelli(1999)}]{jimenez1999}
Jim{\'e}nez J, Pinelli A (1999) The autonomous cycle of near-wall turbulence. J Fluid Mech 389:335--359. \doi{10.1017/S0022112099005066}

\bibitem[{Jung et~al.(1992)Jung, Mangiavacchi, and Akhavan}]{jung1992}
Jung WJ, Mangiavacchi N, Akhavan R (1992) Suppression of turbulence in wall-bounded flows by high-frequency spanwise oscillations. Phys Fluids A 4(8):1605--1607. \doi{10.1063/1.858381}

\bibitem[{K{\"a}hler(2004)}]{kahler2004}
K{\"a}hler CJ (2004) Investigation of the spatio-temporal flow structure in the buffer region of a turbulent boundary layer by means of multiplane stereo {PIV}. Exp Fluids 36:114--130. \doi{10.1007/s00348-003-0680-3}

\bibitem[{K{\"a}hler et~al.(2012)K{\"a}hler, Scharnowski, and Cierpka}]{kahler2012}
K{\"a}hler CJ, Scharnowski S, Cierpka C (2012) On the uncertainty of digital {PIV} and {PTV} near walls. Exp Fluids 52:1641--1656. \doi{10.1007/s00348-012-1307-3}

\bibitem[{Kempaiah et~al.(2020)Kempaiah, Scarano, Elsinga, van Oudheusden, and Bermel}]{kempaiah2020}
Kempaiah KU, Scarano F, Elsinga GE, et~al (2020) 3-dimensional particle image velocimetry based evaluation of turbulent skin-friction reduction by spanwise wall oscillation. Phys Fluids 32:085111. \doi{10.1063/5.0015359}

\bibitem[{Kim(2011)}]{kim2011}
Kim J (2011) Physics and control of wall turbulence for drag reduction. Philos Trans R Soc A 369:1396--1411. \doi{10.1098/rsta.2010.0360}

\bibitem[{Laadhari et~al.(1994)Laadhari, Skandaji, and Morel}]{laadhari1994}
Laadhari F, Skandaji L, Morel R (1994) Turbulence reduction in a boundary layer by a local spanwise oscillating surface. Phys Fluids 6:3218--3220. \doi{10.1063/1.868052}

\bibitem[{Lardeau and Leschziner(2013)}]{lardeau2013}
Lardeau S, Leschziner MA (2013) The streamwise drag-reduction response of a boundary layer subjected to a sudden imposition of transverse oscillatory wall motion. Phys Fluids 25:075109. \doi{10.1063/1.4816290}

\bibitem[{Lima~Pereira et~al.(2020)Lima~Pereira, Ragni, Avallone, and Scarano}]{limapereira2020}
Lima~Pereira LT, Ragni D, Avallone F, et~al (2020) Pressure fluctuations from large-scale {PIV} over a serrated trailing edge. Exp Fluids 61:71. \doi{10.1007/s00348-020-2888-x}

\bibitem[{Orlandi and Jim{\'e}nez(1994)}]{orlandi1994}
Orlandi P, Jim{\'e}nez J (1994) On the generation of turbulent wall friction. Phys Fluids 6:634--641. \doi{10.1063/1.868303}

\bibitem[{Quadrio and Ricco(2004)}]{quadrio2004}
Quadrio M, Ricco P (2004) Critical assessment of turbulent drag reduction through spanwise wall oscillations. J Fluid Mech 521:251--271. \doi{10.1017/S0022112004001855}

\bibitem[{Quadrio et~al.(2009)Quadrio, Ricco, and Viotti}]{quadrio2009}
Quadrio M, Ricco P, Viotti C (2009) Streamwise-travelling waves of spanwise wall velocity for turbulent drag reduction. J Fluid Mech 627:161--178. \doi{10.1017/S0022112009006077}

\bibitem[{Ricco(2004)}]{ricco2004}
Ricco P (2004) Modification of near-wall turbulence due to spanwise wall oscillations. J Turbul 5:N24. \doi{10.1088/1468-5248/5/1/024}

\bibitem[{Ricco and Quadrio(2008)}]{ricco2008}
Ricco P, Quadrio M (2008) Wall-oscillation conditions for drag reduction in turbulent channel flow. Int J Heat Fluid Flow 29:891--902. \doi{10.1016/j.ijheatfluidflow.2007.12.005}

\bibitem[{Ricco and Wu(2004)}]{riccowu2004}
Ricco P, Wu S (2004) On the effects of lateral wall oscillations on a turbulent boundary layer. Exp Therm Fluid Sci 29:41--52. \doi{10.1016/j.expthermflusci.2004.01.010}

\bibitem[{Ricco et~al.(2021)Ricco, Skote, and Leschziner}]{ricco2021}
Ricco P, Skote M, Leschziner MA (2021) A review of turbulent skin-friction drag reduction by near-wall transverse forcing. Prog Aerosp Sci 123:100713. \doi{10.1016/j.paerosci.2021.100713}

\bibitem[{Sanmiguel~Vila et~al.(2017)Sanmiguel~Vila, Vinuesa, Discetti, Ianiro, Schlatter, and {\"O}rl{\"u}}]{sanmiguelvila2017}
Sanmiguel~Vila C, Vinuesa R, Discetti S, et~al (2017) On the identification of well-behaved turbulent boundary layers. J Fluid Mech 822:109--138. \doi{10.1017/jfm.2017.258}

\bibitem[{Scarano(2002)}]{scarano2002}
Scarano F (2002) Iterative image deformation methods in {PIV}. Meas Sci Technol 13:R1--R19. \doi{10.1088/0957-0233/13/1/201}

\bibitem[{Scarano(2003)}]{scarano2003}
Scarano F (2003) Theory of non-isotropic spatial resolution in {PIV}. Exp Fluids 35:268--277. \doi{10.1007/s00348-003-0655-4}

\bibitem[{Scarano and Riethmuller(2000)}]{scarano2000}
Scarano F, Riethmuller ML (2000) Advances in iterative multigrid {PIV} image processing. Exp Fluids 29:S051--S060. \doi{10.1007/s003480070007}

\bibitem[{Schlatter and {\"O}rl{\"u}(2010)}]{schlatter2010}
Schlatter P, {\"O}rl{\"u} R (2010) Assessment of direct numerical simulation data of turbulent boundary layers. J Fluid Mech 659:116--126. \doi{10.1017/S0022112010003113}

\bibitem[{Seber and Wild(1989)}]{seber1989}
Seber GAF, Wild CJ (1989) Nonlinear Regression. Wiley-Interscience, Hoboken, NJ, \doi{10.1002/0471725315}

\bibitem[{Skote(2012)}]{skote2012}
Skote M (2012) Temporal and spatial transients in turbulent boundary layer flow over an oscillating wall. Int J Heat Fluid Flow 38:1--12. \doi{10.1016/j.ijheatfluidflow.2012.08.004}

\bibitem[{Skote et~al.(2019)Skote, Mishra, and Wu}]{skote2019}
Skote M, Mishra M, Wu Y (2019) Wall oscillation induced drag reduction zone in a turbulent boundary layer. Flow Turbul Combust 102:641--666. \doi{10.1007/s10494-018-9979-2}

\bibitem[{Touber and Leschziner(2012)}]{touber2012}
Touber E, Leschziner MA (2012) Near-wall streak modification by spanwise oscillatory wall motion and drag-reduction mechanisms. J Fluid Mech 693:150--200. \doi{10.1017/jfm.2011.507}

\bibitem[{Trujillo et~al.(1997)Trujillo, Bogard, and Ball}]{trujillo1997}
Trujillo S, Bogard D, Ball K (1997) Turbulent boundary layer drag reduction using an oscillating wall. AIAA Paper 97-1870, American Institute of Aeronautics and Astronautics, Snowmass Village, CO, USA. \doi{10.2514/6.1997-1870}

\bibitem[{Wallace(2016)}]{wallace2016}
Wallace JM (2016) Quadrant analysis in turbulence research: History and evolution. Annu Rev Fluid Mech 48:131--158. \doi{10.1146/annurev-fluid-122414-034550}

\bibitem[{Westerweel(1997)}]{westerweel1997}
Westerweel J (1997) Fundamentals of digital particle image velocimetry. Meas Sci Technol 8:1379--1392. \doi{10.1088/0957-0233/8/12/002}

\bibitem[{Westerweel and Scarano(2005)}]{westerweel2005}
Westerweel J, Scarano F (2005) Universal outlier detection for {PIV} data. Exp Fluids 39:1096--1100. \doi{10.1007/s00348-005-0016-6}

\bibitem[{Yakeno et~al.(2014)Yakeno, Hasegawa, and Kasagi}]{yakeno2014}
Yakeno A, Hasegawa Y, Kasagi N (2014) Modification of quasi-streamwise vortical structure in a drag-reduced turbulent channel flow with spanwise wall oscillation. Phys Fluids 26:085109. \doi{10.1063/1.4893903}

\bibitem[{Yudhistira and Skote(2011)}]{yudhistira2011}
Yudhistira I, Skote M (2011) Direct numerical simulation of a turbulent boundary layer over an oscillating wall. J Turbul 12:N9. \doi{10.1080/14685248.2010.538397}

\bibitem[{Zhou et~al.(2024)Zhou, Wang, Cao, and Xu}]{zhou2024}
Zhou L, Wang D, Cao B, et~al (2024) Accurate method for estimating wall-friction based on analytical wall-law model. Aerospace 11:544. \doi{10.3390/aerospace11070544}

\end{thebibliography}

\end{document}